\documentclass[12pt, journal, onecolumn]{IEEEtran}
\usepackage[top = 1in,bottom=1.6in,left=1in,right=1in]{geometry}
\usepackage{titlesec}
\usepackage{setspace}
\usepackage{array}
\usepackage{graphicx}
\usepackage{amssymb}
\usepackage{xcolor}
\usepackage{listings}
\usepackage[english]{babel}
\usepackage{url}
\usepackage{makeidx}
\usepackage{graphicx}
\usepackage{lmodern}
\usepackage{url}
\usepackage{verbatim}
\usepackage{amsmath}
\usepackage{amsfonts}
\usepackage{longtable}
\usepackage{float} 

\newtheorem{theorem}{Theorem}

\newtheorem{definition}{Definition}
\newtheorem{proposition}{Proposition}

\definecolor{mGreen}{rgb}{0,0.6,0}
\definecolor{mGray}{rgb}{0.5,0.5,0.5}
\definecolor{mPurple}{rgb}{0.58,0,0.82}
\definecolor{backgroundColour}{rgb}{0.95,0.95,0.92}

\lstdefinestyle{CStyle}{
    backgroundcolor=\color{backgroundColour},   
    commentstyle=\color{mGreen},
    keywordstyle=\color{magenta},
    numberstyle=\tiny\color{mGray},
    stringstyle=\color{mPurple},
    basicstyle=\footnotesize,
    breakatwhitespace=false,         
    breaklines=true,                 
    captionpos=b,                    
    keepspaces=true,                 
    numbers=left,                    
    numbersep=5pt,                  
    showspaces=false,                
    showstringspaces=false,
    showtabs=false,                  
    tabsize=2,
    language=C
}

\begin{document}

\title{XOR-Based Codes for Private Information Retrieval with Private Side Information}

\author{Murali Krishnan K. H. and J. Harshan\\
Indian Institute of Technology Delhi, India.
}


\IEEEtitleabstractindextext{%
\begin{abstract}
We consider the problem of Private Information Retrieval with Private Side Information (PIR-PSI), wherein a user wants to retrieve a file from replication based non-colluding databases by using the prior knowledge of a subset of the files stored on the databases. The PIR-PSI framework ensures that the privacy of the demand and the side information are jointly preserved, thereby finding potential applications when multiple files have to be downloaded spread across different time-instants. Although the capacity of the PIR-PSI setting is known, we observe that the underlying capacity achieving code construction uses Maximum Distance Separable (MDS) codes thereby contributing to high computational complexity when retrieving the demand. Pointing at this drawback of MDS-based PIR-PSI codes, we propose XOR-based PIR-PSI codes for a simple yet non-trivial setting of two non-colluding databases and two side information files at the user. While our codes offer substantial reduction in complexity when compared to MDS based codes, the code-rate marginally falls short of the capacity of the PIR-PSI setting. Nevertheless, we show that our code-rate is strictly higher than that of XOR-based codes for PIR with no side information, thereby implying that our codes can be useful when downloading multiple files in a sequential manner, instead of applying XOR-based PIR codes on each file.
\end{abstract} 

}

\maketitle

\IEEEdisplaynontitleabstractindextext

%
\IEEEpeerreviewmaketitle

\section{Introduction}
\label{sec:intro}

Private Information Retrieval (PIR) deals with the design of queries to a database so as to provide privacy to the messages downloaded by the user. A trivial way of achieving information-theoretic privacy is to download all the messages stored in the database so that the identity of the demand, i.e., the message the user wants, will be unknown. However, it is well known that this increases the download cost substantially. Ever since the problem of PIR was first introduced in \cite{chor}, various methods have been introduced to efficiently retrieve the demand with privacy, including the code constructions that achieve the capacity of the classic PIR problem \cite{sunjafar}.

Since the contribution of \cite{sunjafar}, numerous variants of the PIR problem have gained attention \cite{banawan}-\cite{sunjafar2}. Among them, an important variant is the problem of PIR with side information. In this setting, the user already knows one or more messages in the database, and she uses this information to reduce the download cost compared to that without side information. The work on PIR with side information began with the cache-aided PIR in \cite{tandon}. Other crucial works on PIR-SI were followed in \cite{Kazemi}-\cite{shariat}. A prominent variant of PIR with side information is the problem of PIR with private side information (PIR-PSI), wherein the privacy of both the demand and the side information are jointly preserved, i.e., the database should not know which message is being queried by the user and which side information is available at her. The application of PIR-PSI has practical significance when users query multiple messages privately from a database. Suppose a user wants to query three messages $A,B,C,$ sequentially with full privacy. One way is to query these three messages separately using the capacity-achieving fully private scheme \cite{sunjafar}. Alternatively, after retrieving $A$ using \cite{sunjafar}, the user can use the prior information on $A$ to download $B$ with a reduced download cost compared to downloading $B$ without side information. Subsequently, with the side information on $A$ and $B$, she can further reduce the download cost for retrieving $C$. This way, as the number of side information messages increases, the download cost for a particular demand can potentially decrease. For these applications, PIR-PSI protocol ensures that demand is downloaded with the help of side information, while jointly preserving the privacy of both the demand and the side information. 

A solution for PIR-PSI was first developed by Kadhe et al. in \cite{kadhe} for the single database setting. Chen et al. further extended it in \cite{zchen} for multiple databases in colluding and non-colluding environments. Both their schemes achieved capacity, however, both the schemes for single database and multiple databases mentioned in \cite{kadhe} and \cite{zchen}, respectively, rely on Maximum Distance Separable (MDS) codes. From the viewpoint of implementation, it is well known that any MDS-coded scheme would have high computational complexity compared to counterpart code that is constructed using XOR bit additions. For example, consider the code for two non-colluding databases with three messages $A,B,C,$ with a length of eight bits per message. Let $A$ be the demand and $C$ be the side information for a particular user. As mentioned in \cite[Section 4.1.1]{zchen}, the query would initially be in the form of Sun-Jafar's capacity-achieving scheme mentioned in \cite{sunjafar}, with four bits of $A$ retrieved out of seven bits queried from a database. This query of seven bits would be converted to a $(7, 13)$ systematic MDS code, e.g., a Reed-Solomon code, and the six bits of the non-systematic part would be downloaded. With the help of the side information $C$, the user can retrieve four demand bits from these six bits, thereby improving the rate. Here, the encoding operation for Reed Solomon code would require $91$ finite field multiplications and $78$ finite field additions, which in turn are computation-intensive tasks compared to executing four XOR bit additions in Sun-Jafar's capacity-achieving scheme. In general, any MDS code for PIR-PSI would depend on the message length $L$, as seen above, and $L$ itself depends on $N^K$ (where $N$ is the number of databases and $K$ is the number of messages). Therefore, the computational complexity will exponentially increase as the number of messages increases.

Capacity-achieving techniques mentioned in \cite{kadhe} and \cite{zchen} are essential to the PIR-PSI problem, however, from the viewpoint of practicality, for a large number of messages, the number of computations would get prohibitively large to create a huge latency in downloading the demand. Therefore, existing PIR-PSI codes are not amenable to implementation in applications that demand low-latency downloads. On the other hand, using a fully private low complexity scheme like \cite{sunjafar} would not be a good idea as this would affect the rate for not exploiting the side information. This implies that one needs to think about constructing PIR-PSI codes based only on XOR computations, and yet providing rates strictly more than \cite{sunjafar}, preferably achieving the capacity of the PIR-PSI problem. 

\subsection{Contributions}
\label{sec:contri}

Inspired by the problem statement discussed above, we make the following contributions in this paper:
\begin{itemize}
    \item We present the first XOR-based code construction for the PIR-PSI problem. Our code construction is applicable for the scenario when $K$ messages are replicated across $N = 2$ non-colluding databases, and the user wishes to retrieve a message with $M = 2$ side information at her side.
    \item Owing to the use of XOR-based queries, we show that our codes offer substantial reduction in the decoding complexity compared to the MDS based counterparts of \cite{zchen}. However, the rate of our codes marginally fall short of the capacity of the PIR-PSI problem for the setting of $N = 2$ non-colluding databases with $M = 2$ side information at the user. On the other hand, the rates of our codes are strictly higher than that of the fully private codes \cite{sunjafar} thereby advertising themselves as a preferred choice when sequentially downloading multiple messages from the databases. For $N=2$ and $M = 2$, a comparison between existing solutions for PIR-PSI and our solution is captured in Table \ref{tab:my_label}.
    \item For the proposed code construction, we prove the joint privacy property by explicitly showing that the query to one of the databases can be fixed, and then the query to the other database can be modified in such a way that any combination of side information can be used to retrieve any demand from the two databases. 
\end{itemize}
\begin{table}[H]
    \centering
   \begin{tabular}{|c|c|c|c|}
    \hline
         Scheme&Rate&Method&Computational \\
         &&&complexity \\
         \hline
          \cite{zchen}&$\frac{2^{K-3}}{2^{K-2}-1}$&MDS& High. Exponential as \\ 
         & (higher than & coding & $K$ increases. Finite field\\  
         & our code) && multiplication involved.\\
         \hline
        Our&$\frac{2^{K-2}}{2^{K-1}-1}$&XOR&Low. Bit-wise XOR\\ 
        scheme&&&involved\\
        \hline
       \cite{sunjafar}&$\frac{2^{K-1}}{2^{K}-1}$&XOR&Low. Bit-wise XOR involved \\ 
       & (lower than & &\\
       & our code) & &\\
       \hline
        
    \end{tabular}
    \caption{}
    \label{tab:my_label}
\end{table}
 
The rest of the paper concerns the following sections. Section \ref{sec2:problem} presents the problem statement and related notations. Section \ref{sec:3} presents the code construction, whereas Section \ref{sec:4} provides the proof for joint privacy of the XOR-based codes for arbitrary values of $K$. Section \ref{sec:privacy_proof_K_7} exemplifies the privacy proof for the case when $K = 7$. Finally, some directions for future research are presented in Section \ref{sec:5}. 

\section{Problem Statement}
\label{sec2:problem}

Consider two replicated non-colluding databases, $N_1$ and $N_2$ with $K$ messages namely $X_1, X_2, \ldots, X_{K-1}, X_K$. All the $K$ messages are of size $L$ bits which are independent and uniformly distributed. Therefore, we denote the $i$-th file $ X_{i}$ as $X_{i} = [X_{i,1}, X_{i,2}, \ldots, X_{i,L}]$. Among the $K$ messages, let $X_\gamma$, for $\gamma \in [K]$, be the demand message for a user who already has $M = 2$ side information messages $X_\alpha$ and $X_\beta$, such that $\alpha, \beta \in [K]$\textbackslash$\{\gamma\}$ satisfying $\alpha \neq \beta$. In this model the messages other than $X_{\gamma}, X_{\alpha}$ and $X_{\beta}$ are referred to as byproducts. The user wishes to leverage the knowledge of $X_\alpha$ and $X_\beta$ to retrieve $X_\gamma$ by downloading fewer bits from each database compared to retrieving $X_{\gamma}$ using the scheme in \cite{sunjafar} without using the side information. At the same time, the user also wants to jointly preserve the privacy of the demand $X_\gamma$ and the side information messages $X_\alpha$ and $X_\beta$. i.e., the indexes $\gamma$, $\alpha$ and $\beta$ should be unknown to both databases. In the context of this work, the code for retrieving $X_\gamma$ should be such that each database provides the user a set of XOR additions of the bits of the $K$ messages. This way, the user would be able to retrieve $X_\gamma$ by performing simple XOR bit additions on the downloaded bits while using the prior knowledge of $X_\alpha$ and $X_\beta$. 

Towards constructing an XOR-based code for PIR-PSI, we make the following definitions. In order to retrieve the demand $X_{\gamma}$ using the side information $X_{\alpha}$ and $X_{\beta}$, every combination of message bits submitted to a database is referred to as \emph{codeword}, the set of codewords submitted to a database is called a \emph{query}, and the union of queries submitted to $N_{1}$ and $N_{2}$ is referred to as the \emph{code}. When designing the codewords of a query it is important to find the right set of XOR combinations of the $K$ message indexes, and then choose the appropriate bit locations of the message indexes in the XOR combinations. To formally describe the XOR combinations of the message indexes, we introduce the following definitions.

\begin{definition}
\label{def_ingre}
For a finite set $\mathcal{M} = \{M_{1}, M_{2}, \ldots, M_{P}\}$ containing $P$ distinct variables, we define a mapping $\Phi$ such that $\Phi(\mathcal{M}) = \sum_{i = 1}^{P} M_{i}$ if $P \geq 1$, and $\Phi(\mathcal{M}) = \phi$ if $\mathcal{M}$ is empty set.
\end{definition}

\begin{definition}
\label{def_power_set_struct}
With $\mathcal{X} = \{X_1, X_2, \ldots, X_{K}\}$, we define a power set based skeleton structure $P(\mathcal{X}) = \{ \Phi(\mathcal{V}) ~|~ \forall ~\mathcal{V} \in PS(\mathcal{X})\}$, where $\Phi(\cdot)$ is as introduced in Definition \ref{def_ingre}, and $PS(\mathcal{X})$ is the power set of $\mathcal{X}$. 
\end{definition}

From Definition \ref{def_power_set_struct}, it is clear that the queries (without specifying the bit locations of each message)  submitted to $N_{1}$ and $N_{2}$ must be subsets of $P(\mathcal{X})$ \textbackslash $\{\phi\}$, wherein the $+$ operator in $\Phi$ is treated as XOR operation. Henceforth, throughout the paper, the queries submitted to $N_{1}$ and $N_{2}$ to retrieve $X_{\gamma}$ using the side information $X_{\alpha}$ and $X_{\beta}$ are denoted by $\mathcal{C}_{1}$ and $\mathcal{C}_{2}$, respectively, and the overall code is denoted by $(\mathcal{C}_{1}, \mathcal{C}_{2})$. To propose a formal design criteria to choose the subsets of $P(\mathcal{X})$ \textbackslash $\{\phi\}$ as queries, we define a singleton block in a query as the set of codewords that consist of only a single message bit. Similarly, we define an $n$-tuple sum block in a query, for $2 \leq n < K$, as the set of codewords that consist of XOR bit additions of $n$ different messages.

\subsection{Design Criteria for XOR-based PIR-PSI Codes}
\label{subsec:design_criteria}

For $N = 2$, $M = 2$ and any $K \geq 3$, let $(\mathcal{C}_{1}, \mathcal{C}_{2})$ be an XOR-based code to retrieve $X_{\gamma}$ using the side information $X_{\alpha}$ and $X_{\beta}$ The code $(\mathcal{C}_{1}, \mathcal{C}_{2})$ is said to be XOR-based PIR-PSI code if keeping the query to $N_1$ (or $N_{2}$) unchanged, it is possible to design a query to $N_{2}$ (or $N_{1}$) such that 
\begin{enumerate}
    \item \textbf{Condition 1}: any demand can be retrieved from $N_1$ and $N_{2}$ using any two side information messages.
    \item \textbf{Condition 2}: the structure of the new query to $N_2$ (or $N_{1}$) is same as that of $\mathcal{C}_{2}$ (or $\mathcal{C}_{1}$), namely: (i) the number of singleton and $n$-tuple sum blocks is the same, for any $2 \leq n \leq K-1$, and (ii) the frequency distribution of the message bits across all the codewords in the query is the same as that of $\mathcal{C}_2$ (or $\mathcal{C}_{1}$).
\end{enumerate}

We take a two-step approach to design codes satisfying the above criteria. First, we present the queries of the code for a given demand $X_{\gamma}$ and a pair of side information $X_{\alpha}$ and $X_{\beta}$, and prove the correctness of the construction in retrieving $X_{\gamma}$. Subsequently, we present a rigorous proof to show that keeping the query to $N_1$ (or $N_{2}$) unchanged, new queries to $N_2$ (or $N_{1}$) can be constructed by satisfying \textbf{Condition 1} and \textbf{Condition 2}. We show that the rate of code is more than that of \cite{sunjafar}, implying that our codes can be used when sequentially downloading multiple files. The construction procedure for XOR-based PIR-PSI code is explained in the next section.


\section{XOR-Based PIR Codes with Private Side Information}
\label{sec:3}

Among the $K$ messages, let $X_1$ be the demand, and $X_2$ and $X_3$ be the side information. With $M = 2$, our construction is applicable only for $K \geq 3$. For $K = 3$, the code construction is trivial with rate one since an XOR version of all the files can be downloaded. For $K > 3$, the ingredients and the instructions provided in the next two sections must be followed to obtain the queries for $N_1$ and $N_2$. Although the individual bits of the $K$ files will be used to retrieve the $L$ bits of $X_{1}$, first, our construction  provides a way to place the file index in the query, and then describes a way to choose the specific bits of each file in the query. Along with the steps for ingredients and code construction, a running-example for $K=4$, with messages $A,B,C,D$ is also presented, wherein $A$ plays the role of demand, i.e., $X_{1}$, $B$ and $C$ play the role of side information, i.e., $X_2$, and $X_3$. 

\subsection{Ingredients and Construction Strategy}
\label{sec:3i}

With $\mathcal{X} = \{X_1, X_2, \ldots, X_{K-1}\}$, we construct $P(\mathcal{X}) = \{ \Phi(\mathcal{V}) ~|~ \forall ~\mathcal{V} \in PS(\mathcal{X})\}$, where $\Phi(\cdot)$ is as introduced in Definition \ref{def_ingre}, $PS(\mathcal{X})$ is the power set of $\mathcal{X}$. In the context of the running-example with $K = 4$, we have $\mathcal{X} = \{A, B, C\}$, and therefore, $P(\{A, B, C\})$ is given in Table \ref{tab:1}.
    
\begin{table}[h!]
      \centering
\begin{tabular}{ |c| } 
 \hline
 $P(\{A,B,C\})$\\\hline
 $\phi$ \\ 
 A \\ 
 B \\
 C \\
 A+B\\
 A+C\\
 B+C\\
 A+B+C\\
 \hline
\end{tabular}
      \caption{}
      \label{tab:1}
  \end{table}
  
Owing to the use of power set, the elements of  $P(\mathcal{X})$ are unique, generating a total of $2^{K-1}$ elements by using $2^{K-2}$ copies of each message. Towards converting $P(\mathcal{X})$ into a query, we will allocate distinct indexes for each copy of the message, thereby resulting in $2^{K-2}$ unique bits of a message. This will ensure that the query at this stage will contain $K -1$ messages each containing $L = 2^{K-2}$ bits. In order to prepare the desired query with $K$ messages, we need to add $2^{K-2}$ copies of the message $X_{K}$ to the existing elements of $P(\mathcal{X})$. In the context of the example,  as seen in Table \ref{tab:1}, 4 bits of $A,B,C$ are present in $P(\{A,B,C\})$. Therefore, 4 copies of $D$ should be added at different positions. In the next section, we provide a set of instructions to add $X_{K}$. 

\subsection{Algorithm for Code Construction}

  \textbf{1.} Rewrite $P(\{X_{1}, X_{2}, \ldots, X_{K-1}\})$ as $P(\{X_1,X_2\})$ $\bigotimes$  $P(\{X_3,X_4,$ ...$, X_{K-1}\})$, where $\bigotimes$ operator can be defined on two sets $\mathcal{M}_{1}$ and $\mathcal{M}_{2}$ as 
  \begin{equation*}
      \mathcal{M}_{1} \bigotimes \mathcal{M}_{2} = \{\alpha + \beta, ~|~ \forall ~\alpha \in \mathcal{M}_{1}, \beta \in \mathcal{M}_{2}\},
  \end{equation*}
  such that $\phi + \phi = \phi$, $\alpha + \phi = \alpha$ and $\phi + \beta = \beta$. For the example with $K = 4$, it is straightforward to verify that applying $\bigotimes$ operator on $P(\{A,B\})$ and $P(\{C\})$, as given in Table \ref{tab:2}, gives $P(\{A, B, C\})$ given in Table  \ref{tab:1}.

  \begin{table}[h!]
      \centering
\begin{tabular}{ |c|c| } 
\hline
$P(\{A,B\})$ & $P(\{C\})$\\
\hline
 $\phi$& $\phi$ \\ 
A & C\\
B & \\
A+B & \\
 \hline
\end{tabular}
 \caption{}
\label{tab:2}
 \end{table}
 
    
 \textbf{2.} From Step \textbf{1}, we know that the size of $P(\{X_3,X_4,$ ...$, X_{K-1}\})$ is $2^{K-3}$. Form a new table with two columns, namely: Column 1 and Column 2. Excluding $\phi$, replicate the $2^{K-3}-1$ entries of $P(\{X_3,X_4,$ ...$, X_{K-1}\})$ into two columns, thereby forming a total of $2(2^{K-3}-1)=2^{K-2}-2$ elements. For the example, $\phi$ is removed from $P(\{C\})$, and the remaining element is replicated into 2 columns as shown in Table \ref{tab:3}.
  
  \begin{table}[h!]
      \centering
\begin{tabular}{ |c|c| } 
\hline
Column 1 & Column 2\\
\hline
C&C\\
 \hline
\end{tabular}
 \caption{}
      \label{tab:3}
  \end{table}

\textbf{3.} Add $X_K$ to all the entries of Column 1. Leave Column 2 unaltered. Through this step, out of the $2^{K-2}$ copies, $2^{K-3}-1$ copies of $X_K$ are added. For the example, the two columns are as shown in Table \ref{tab:4}.
\begin{table}[h!]
    \centering
   \begin{tabular}{ |c|c| } 
\hline
Column 1 & Column 2\\
\hline
C+D&C\\
 \hline
\end{tabular}
    \caption{}
    \label{tab:4}
\end{table}
  
 \textbf{4.} Form a new table with two columns, namely: Column 1' and Column 2'. Perform $\bigotimes$ operation between $\{\phi,X_1+X_2\}$ and all the entries in the two-tuple sum block and three-tuple sum block of Column 1 from Step \textbf{3}, and place the result in Column 1'. Similarly, perform $\bigotimes$ operation between $\{X_1, X_2\}$ and all the entries in the singleton block and two-tuple sum block of Column 2 from Step \textbf{3}, and place the result in Column 2'.
 In the example, only a singleton block is present, and therefore, the corresponding Column 1' and Column 2' are presented in Table \ref{tab:5}.
 
\begin{table}[H]
    \centering
\begin{tabular}{ |c|c| } 
\hline
Column 1' & Column 2'\\
\hline
C+D&C+A\\
C+D+A+B&C+B\\
 \hline
\end{tabular}
    \caption{}
    \label{tab:5}
\end{table}

 \textbf{5.} Skip this step if $K<=5$ since largest value of $n$ for the $n$-tuple sum block is 2. This is already addressed in the previous step.
 \begin{itemize}
     \item If $K$ = 6 or 7, perform $\bigotimes$ operation between $\{X_1, X_2\}$ and all the entries in the $n$-tuple sum block, for $n \in \{4, 5, \ldots, K-2\}$ of Column 1, and append the result in Column 1'. Similarly, perform $\bigotimes$ operation between $\{\phi, X_1 + X_2\}$ and all the entries in the $n$-tuple sum block, for $n \in \{3, 4, \ldots, K-3\}$, of Column 2, and append the result in Column 2'. 
     
     \item If $K>7$, perform $\bigotimes$ operation between $\{X_1, X_2\}$ and all the entries in the $n$-tuple sum block, for $n \in \{4, 5, \ldots, K-3\}$ of Column 1, and append the result in Column 1'. Similarly, perform $\bigotimes$ operation between $\{\phi, X_1 + X_2\}$ and all the entries in the $n$-tuple sum block, for $n \in \{3, 4, \ldots, K-4\}$, of Column 2, and append the result in Column 2'. However, for the $(K-2)$-tuple sum block in Column 1, perform $\bigotimes$ operation with $\{\phi,X_1+X_2\}$ and append the result in Column 1'. Similarly, for the $(K-3)$-tuple sum block in Column 2, perform $\bigotimes$ operation $\{X_1,X_2\}$ and append the result in Column 2'.
 \end{itemize}
This step is not applicable to the running-example since $K = 4$. At the end of this step, both Column 1' and Column 2' contain $2^{K-2}-2$ elements each owing to the $\bigotimes$ operation. With this, we highlight that the union of Column 1'and Column 2' has generated $2^{K-1}-4$ elements of $P(\{X_{1}, X_{2}, \ldots, X_{K-1}\})$. This does not include $\{\phi, X_{1}, X_{2}, X_{1} + X_{2}\}$ since $\phi$ of $P(\{X_{3}, X_{4}, \ldots, X_{K-1}\})$ was excluded when constructing Column 1 and Column 2 in Step \textbf{2}. Furthermore, since $X_K$ was already added to Column 1 (containing $2^{K-3}-1$ elements), at the end of Step \textbf{5}, Column 1' contains $2^{K-2}-2$ copies of $X_{K}$. This implies that only two more copies of $X_K$ are to be added to ensure that each message has $2^{K-2}$ copies. In the example, Step \textbf{3} added one copy of $D$, and Step \textbf{4} made it two copies. Now two more copies are remaining.
    
\textbf{6.} From $\{\phi,X_1,X_2,X_1+X_2\}$, omit $\phi$, and add $X_K$ to $X_2$. This generates $\{X_1,X_2+X_K,X_1+X_2\}$. Place these elements in a new column, referred to as Column 3. Note that one more copy of $X_{K}$ must be added to achieve $2^{K-2}$ copies. In the example, Column 3 is given in Table \ref{tab:6}. 
\begin{table}[H]
    \centering
   \begin{tabular}{ |c| } 
   \hline
   Column 3\\
   \hline
   A\\
   B+D\\
   A+B\\
   \hline
   \end{tabular}
    \caption{}
    \label{tab:6}
\end{table}
    
    \textbf{7.} Form the query to $N_{1}$ by taking the union of all the elements in Column 1', Column 2' and Column 3. By construction, this set contains $X_1+X_3$ coming from Column 2'. Therefore, add the last remaining copy of $X_K$ to $X_1+X_3$, and update it as $X_1+X_3 + X_{K}$. Thus, we have a total of $2^{K-1}-1$ elements in this query constructed by adding $2^{K-2}$ copies of $K$ files $\{X_{1}, X_{2}, \ldots, X_{K}\}$. Finally, in the query to $N_{1}$, provide distinct indexes to every copy of a message thereby ensuring that every bit of a message is used only once. Overall, the query is a set of linear combinations of $\{X_{i, j} ~|~ i \in [K], j \in [L]\}$. In the example, the query to $N_{1}$ after following the above steps are given in Table \ref{tab:8_without_i} (without indexes) and Table \ref{tab:8} (with indexes). 
    \begin{figure}[H]
  \begin{minipage}{.5\textwidth}  
    \begin{table}[H]
        \centering
      \begin{tabular}{ |c| } 
   \hline
QUERY TO $N_1$ WITHOUT INDEXES\\
   \hline
   $A$\\
   $A + B$\\
   $B + C$\\
   $B + D$\\
   $C + D$\\
   $A + C + D$\\
   $A + B + C + D$\\
   \hline
   \end{tabular}
        \caption{}
        \label{tab:8_without_i}
    \end{table}
    \end{minipage}
    \begin{minipage}{.5\textwidth}  
    \begin{table}[H]
        \centering
      \begin{tabular}{ |c| } 
   \hline
QUERY TO $N_1$ WITH INDEXES\\
   \hline
   $A_1$\\
   $A_2+B_1$\\
   $B_2+C_1$\\
   $B_3+D_1$\\
   $C_2+D_2$\\
   $A_3+C_3+D_3$\\
   $A_4+B_4+C_4+D_4$\\
   \hline
   \end{tabular}
        \caption{}
        \label{tab:8}
    \end{table}
    \end{minipage}
    \end{figure}
    
Before we present the procedure to construct the query to $N_{2}$, we present special structures in the query defined as \emph{the known byproduct combination} and the \emph{unknown byproduct combination}, and then present some results on their structure. Known byproduct combination are the bit combinations of byproducts in a codeword that does not contain the demand index. Formally, if $X_{1}$ is the demand, $X_{2}, X_{3}$ are the side information, and the format of the codeword is $H + W$, where $H \in P(\{X_{2}, X_{3}\})$ and $W \in P(\{X_{4}, X_{5}, \ldots, X_{K}\})$\textbackslash $\{\phi\}$, then $W$ is the known byproduct combination. Informally, this is the combination of byproduct messages that can be retrieved from the database. Unknown byproduct combinations are the bit combinations of the byproducts in a codeword that contains the demand index with or without the side information message bits. Formally, if $X_{1}$ is the demand, $X_{2}, X_{3}$ are the side information, and the codeword is of the form $X_{1} + H + W$, where $H \in P(\{X_{2}, X_{3}\})$ and $W \in P(\{X_{3}, X_{4}, \ldots, X_{K}\})$ \textbackslash $\{\phi\}$, then $W$ is the unknown byproduct combination. Informally, this is the combination of byproduct messages that cannot be retrieved from the database due to the unknown demand index that is along with it.
    
\begin{proposition}
\label{prop:symm_known_unknown}
If the message combination $W \in P(\{X_{4}, X_{5}, \ldots, X_{K}\})$\textbackslash$\{\phi\}$ appears as an unknown byproduct in the query to $N_{1}$, then it also exists as a known byproduct, however, with different index values on each message.
\end{proposition}
\begin{IEEEproof}
By definition, if $W$ is an unknown byproduct, then it appears in the query along with the demand $X_{1}$. Since $W$ may appear alone or along with the side information messages, we shall denote the query associated with the unknown byproduct as $U = X_{1} + f(X_{2}, X_{3}) + W$, where  $f(X_{2}, X_{3}) \in \{\phi, X_{2}, X_{3}, X_{2} +X_{3}\}$. From the code construction, $U$ must be equal to $X_{1} + X_{3} + X_{K}$ that was added in Step \textbf{7}, or it must belong to either Column 1' or Column 2' of Step \textbf{4} and Step \textbf{5}. If $U = X_{1} + X_{3} + X_{K}$, then the corresponding known byproduct is available in Column 3 of Step \textbf{6}. However, if $U$ is available in Column 1', then the corresponding known byproduct is also in Column 1' because the elements of Column 1' are generated by performing $\bigotimes$ operation with either $\{X_{1}, X_{2}\}$ or $\{\phi, X_{1} + X_{2}\}$. The same argument is also applicable if $U$ is available in Column 2'. Finally, the index values used on the known byproduct are different from that of unknown byproduct since every copy of a message is assigned different indexes as per Step \textbf{7}. This completes the proof. \\  
\end{IEEEproof}   

\noindent \textbf{8.} To generate query to $N_{2}$, the following instructions must be followed. Copy the structure of $N_1$ as it is without the indexes of bits. For the demand $X_{1}$, the first $2^{K-2}$ bits are already queried in $N_1$. Therefore, give the next $2^{K-2}$ numbers as the indexes of $X_1$ in $N_2$. Use the same index number on each message of the side information as that of the query in $N_{1}$. From Proposition \ref{prop:symm_known_unknown}, the query to $N_1$ produces a symmetric sequence of known and unknown byproduct combinations. List the unknown and known byproduct combinations in two separate columns in the ascending order of $n$-tuple length following the lexicographical order and ascending order of indexes of bits. For a given byproduct combination of messages in the unknown column, an identical byproduct combination exists in the known column, however, with the difference that the index values used by the unknown combination is different from the known combination. To assign index values on the byproduct messages of $N_{2}$, for a given byproduct combination, use the index values of the unknown combination of $N_{1}$ on the known combination of $N_{2}$, and vice-versa. This ensures that all the byproduct messages can be indexed using one-to-one mapping between the unknown and the known column.

For the given example, in the fourth and fifth bit of Table \ref{tab:8}, $D_1$ and $D_2$ are with side information, and therefore, they become the known byproduct bits. Similarly, in the sixth and seventh bit of the query to $N_1$, $D_3$ and $D_4$ are with the demand, and therefore, they become unknown byproduct bits, as listed in Table \ref{tab:9}. To generate the query for $N_2$, the structure of $N_1$ is copied without the indexes. The indexes of the demand are $\{5, 6, 7, 8\}$ in $N_2$. The indexes of $B,C$ are maintained as they are. Based on the one-to-one mapping in Table \ref{tab:9}, $D_{3}$ and $D_{1}$ are swapped, and so are $D_{4}$ and $D_2$. Finally, the query for $N_{2}$ is as shown in Table \ref{tab:10}. This completes the code construction. 

\begin{figure}[H]
  \begin{minipage}{.5\textwidth}  
\begin{table}[H]
    \centering
   \begin{tabular}{ |c|c| } 
\hline
UNKNOWN&KNOWN\\
\hline
$D_3$&$D_1$\\
$D_4$&$D_2$\\
\hline
\end{tabular}
    \caption{}
    \label{tab:9}
\end{table}
\end{minipage}
\begin{minipage}{.5\textwidth} 
\begin{table}[H]
    \centering
    \begin{tabular}{ |c|c| } 
   \hline
QUERY TO $N_1$&QUERY TO $N_2$\\
   \hline
   $A_1$&$A_5$\\
   $A_2+B_1$&$A_6+B_1$\\
   $B_2+C_1$& $B_2+C_1$\\
   $B_3+D_1$&$B_3+D_3$\\
   $C_2+D_2$&$C_2+D_4$\\
   $A_3+C_3+D_3$&$A_7+C_3+D_1$\\
   $A_4+B_4+C_4+D_4$&$A_8+B_4+C_4+D_2$\\
   \hline
   \end{tabular}
    \caption{}
    \label{tab:10}
\end{table}
\end{minipage}
\end{figure}

\begin{theorem}
For $N = 2$ and any $K > 3$, with the knowledge of side information $X_2$ and $X_3$, the proposed code construction can retrieve $2^{K-2}$ bits of $X_1$ per database by downloading $2^{K-1} - 1$ bits per database. Thus, the rate of the code is
\begin{equation}
\label{eq:rate}
R = \frac{2^{K-2}}{2^{K-1}-1}.
\end{equation}
\end{theorem}
\begin{IEEEproof}
From the code construction, the query to $N_{1}$ and $N_{2}$ are obtained by adding $2^{K-2}$ bits of $X_{K}$ to the power set structure $P(\{X_{1}, X_{2}, \ldots, X_{K-1}\})$\textbackslash$\{\phi\}$. Since the cardinality of $P(\{X_{1}, X_{2}, \ldots, X_{K-1}\})$\textbackslash$\{\phi\}$ is $2^{K-1} - 1$, and every message in $P(\{X_{1}, X_{2}, \ldots, X_{K-1}\})$\textbackslash$\{\phi\}$ appears $2^{K-2}$ times, the rate of the code is as given in \eqref{eq:rate}. In the rest of the proof, we show that every bit of $X_{1}$ can be retrieved from $N_{1}$ and $N_{2}$ using the side information. If a bit of $X_{1}$ appears in the form of $X_{1} + f(X_{2}, X_{3})$ either in $N_{1}$ or $N_{2}$, where $f(X_{2}, X_{3}) \in \{\phi, X_{2}, X_{3}, X_{2} +X_{3}\}$, then this bit can be retrieved since $X_{1}$ and $X_{2}$ are known. On the other hand, if a bit of $X_{1}$ appears in the form of $X_{1} + f(X_{2}, X_{3}) + W$ on database $N_{1}$ (or $N_{2}$), where $W \in P(\{X_{4}, X_{5}, \ldots, X_{K}\})$\textbackslash$\{\phi\}$, then from Step \textbf{8} of the code construction this bit can also be retrieved by using the side information and the corresponding known byproduct of $W$, which is downloaded from $N_{2}$ (or $N_{1}$). This completes the proof. 
\end{IEEEproof}

Although we provided a running-example for construction with $K = 4$, we present another example with $K = 7$ in the next section. The example for $K = 7$ required later for the purpose of proving joint privacy.

\subsection{Example for K=7}
\label{sec:3a}

Consider 7 messages $A,B,C,D,E,F,G$. Let $A$ be the demand and $B$ and $C$ be the side information known, and $D,E,F$ and $G$ are the byproducts. The step-by-step construction of query to retrieve $A$ from $N_1$ and $N_2$ is shown below. Construct the skeleton power set structure $P(\{A,B,C,D,E,F\})$ for $K-1=6$ messages. A total of $2^{K-1}=2^{7-1}=64$ elements are formed.
 
 \begin{center}
\begin{longtable}{ |c| } 
 \hline
$P(\{A,B,C,D,E,F\})$\\\hline
 $\phi$ \\ 
A\\
B\\
C\\
D\\
E\\
F\\
A+B\\
A+C\\
A+D\\
A+E\\
A+F\\
B+C\\
B+D\\
B+E\\
B+F\\
C+D\\
C+E\\
C+F\\
D+E\\

D+F\\
E+F\\
A+B+C\\
A+B+D\\
A+B+E\\
A+B+F\\
A+C+D\\
A+C+E\\
A+C+F\\
A+D+E\\
A+D+F\\
A+E+F\\
B+C+D\\
B+C+E\\
B+C+F\\
B+D+E\\
B+D+F\\
B+E+F\\
C+D+E\\
C+D+F\\
C+E+F\\
D+E+F\\
A+B+C+D\\
A+B+C+E\\
A+B+C+F\\
A+B+D+E\\
A+B+D+F\\
A+B+E+F\\
A+C+D+E\\
A+C+D+F\\
A+C+E+F\\
A+D+E+F\\
B+C+D+E\\
B+C+D+F\\
B+C+E+F\\
B+D+E+F\\

C+D+E+F\\
A+B+C+D+E\\
A+B+C+D+F\\
A+B+C+E+F\\
A+B+D+E+F\\

A+C+D+E+F\\
B+C+D+E+F\\
A+B+C+D+E+F\\
 \hline
 \caption{Power set structure with $K = 7$}
 \label{tab:11}
\end{longtable}

\end{center}
With the message $G$ playing the role of $X_{K}$ as per the construction in Section \ref{sec:3}, following are the steps to add 32 copies of $G$ to the existing elements of $P(\{A,B,C,D,E,F\})$.

 \begin{enumerate}
\item  Rewrite $P(\{A,B,C,D,E,F\})$ in the form $P(\{$A$,$B$\})$ $\bigotimes$  $P(\{$C,D,E,F$\})$.

\begin{center}
\begin{longtable}{ |c|c| } 
\hline
$P(\{$A$,$B$\})$&$P(\{$C,D,E,F$\})$\\\hline
 $\phi$&$\phi$ \\ 
A&C\\
B&D\\
A+B&E\\
&F\\
&C+D\\
&C+E\\
&C+F\\
&D+E\\
&D+F\\
&E+F\\
&C+D+E\\
&C+D+F\\
&C+E+F\\
&D+E+F\\
&C+D+E+F\\
 \hline
 \caption{$P(\{$A$,$B$\})$ $\bigotimes$  $P(\{$C,D,E,F$\})$}
 \label{tab:12}

\end{longtable}
\end{center}
\item Now remove $\phi$ from $P(\{$C,D,E,F$\})$ and replicate the remaining $2^{K-3}-1=15$ bits of $P(\{$C,D,E,F$\})$\textbackslash$\{\phi\}$ into 2 columns namely: Column 1 and Column 2, forming a total of $2(2^{K-3}-1)=30$ bits.
\newpage
\begin{center}
\begin{longtable}{ |c|c| } 
\hline
Column 1& Column 2\\
\hline
C&C\\

D&D\\
E&E\\
F&F\\
C+D&C+D\\
C+E&C+E\\

C+F&C+F\\
D+E&D+E\\
D+F&D+F\\
E+F&E+F\\
C+D+E&C+D+E\\
C+D+F&C+D+F\\
C+E+F&C+E+F\\
D+E+F&D+E+F\\
C+D+E+F&C+D+E+F\\
 \hline
 \caption{Outcome of Step 2 with $K = 7$}
 \label{tab:13}

\end{longtable}
\end{center}

\item Add $G$ to all the entries of Column 1. Leave Column 2 unaltered.
\begin{center}
\begin{longtable}{ |c|c| } 
\hline
Column 1& Column 2\\
\hline
C+G&C\\
D+G&D\\
E+G&E\\
F+G&F\\
C+D+G&C+D\\
C+E+G&C+E\\
C+F+G&C+F\\
D+E+G&D+E\\
D+F+G&D+F\\
E+F+G&E+F\\
C+D+E+G&C+D+E\\
C+D+F+G&C+D+F\\
C+E+F+G&C+E+F\\
D+E+F+G&D+E+F\\
C+D+E+F+G&C+D+E+F\\
 \hline
 \caption{Outcome of Step 3 with $K = 7$}
 \label{tab:14}

\end{longtable}
\end{center}
\item  Perform $\bigotimes$ operation between $\{\phi,X_1+X_2\}=\{\phi,A+B\}$ and all the entries in the two-tuple sum block and the three-tuple sum block of Column 1 and place the result in Column 1'. Similarly, perform $\bigotimes$ operation between $\{X_1,X_2\}=\{A,B\}$ and all the entries in the singleton block and two-tuple sum block of Column 2 and place the result in Column 2'.  
\begin{center}
\begin{longtable}{ |c|c| } 
\hline
Column 1' & Column 2'\\
\hline
C+G&C+A\\
C+G+A+B&C+B\\
D+G&D+A\\
D+G+A+B&D+B\\
E+G&E+A\\
E+G+A+B&E+B\\
F+G&F+A\\
F+G+A+B&F+B\\
C+D+G&C+D+A\\
C+D+G+A+B&C+D+B\\
C+E+G&C+E+A\\
C+E+G+A+B&C+E+B\\
C+F+G&C+F+A\\
C+F+G+A+B&C+F+B\\
D+E+G&D+E+A\\
D+E+G+A+B&D+E+B\\
D+F+G&D+F+A\\
D+F+G+A+B&D+F+B\\
E+F+G&E+F+A\\
E+F+G+A+B&E+F+B\\
C+D+E+G&C+D+E\\
C+D+F+G&C+D+F\\
C+E+F+G&C+E+F\\
D+E+F+G&D+E+F\\
C+D+E+F+G&C+D+E+F\\
 \hline
 \caption{Outcome of Step 4 with $K = 7$}
 \label{tab:15}

\end{longtable}
\end{center}
\item Since $K=7$ in this case, perform $\bigotimes$ operation between $\{A,B\}$ and all the entries in the $n$-tuple sum block, for $n \in \{4, 5, \ldots, K-2\}$ of Column 1, and append the result in Column 1'. Similarly, perform $\bigotimes$ operation between $\{\phi,A+B\}$ and all the entries in the $n$-tuple sum block, for $n \in \{3, 4, \ldots, K-3\}$, of Column 2, and append the result in Column 2'.
\begin{center}
\begin{longtable}{ |c|c| } 
\hline
Column 1' & Column 2'\\
\hline
C+G&C+A\\
C+G+A+B&C+B\\
D+G&D+A\\

D+G+A+B&D+B\\
E+G&E+A\\
E+G+A+B&E+B\\
F+G&F+A\\
F+G+A+B&F+B\\
C+D+G&C+D+A\\
C+D+G+A+B&C+D+B\\
C+E+G&C+E+A\\
C+E+G+A+B&C+E+B\\
C+F+G&C+F+A\\

C+F+G+A+B&C+F+B\\
D+E+G&D+E+A\\
D+E+G+A+B&D+E+B\\
D+F+G&D+F+A\\
D+F+G+A+B&D+F+B\\
E+F+G&E+F+A\\
E+F+G+A+B&E+F+B\\
C+D+E+G+A&C+D+E\\
C+D+E+G+B&C+D+E+A+B\\
C+D+F+G+A&C+D+F\\
C+D+F+G+B&C+D+F+A+B\\
C+E+F+G+A&C+E+F\\
C+E+F+G+B&C+E+F+A+B\\
D+E+F+G+A&D+E+F\\
D+E+F+G+B&D+E+F+A+B\\
C+D+E+F+G+A&C+D+E+F\\
C+D+E+F+G+B&C+D+E+F+A+B\\
 \hline
 \caption{Outcome of Step 5 with $K = 7$}
 \label{tab:16}

\end{longtable}
\end{center}
\item From $\{\phi,A,B,A+B\}$, omit $\phi$, and add $G$ to $B$. This generates $\{A,B+G,A+B\}$. Place these elements in a new column, referred to as Column 3.\\
\begin{table}[H]
    \centering
   \begin{tabular}{ |c| } 
   \hline
   Column 3\\
   \hline
   A\\
   B+G\\
   A+B\\
   \hline
   \end{tabular}
    \caption{Outcome of Step 6 with $K = 7$}
    \label{tab:17}
\end{table}

\item  Form the query to the database $N_{1}$ by taking the union of all the elements in Column 1', Column 2' and Column 3. Add $G$ to $A+C$, and update it as $A+C+G$. In the query to $N_{1}$, provide distinct indexes to every copy of a message as seen below. Arranging the query in lexicographical order, we get the following query. 
      \begin{center}
\begin{longtable}{ |c| } 
\hline
QUERY TO $N_1$\\
\hline
$A_1$\\
$A_2+B_1$\\
$A_3+D_1$\\
$A_4+E_1$\\
$A_5+F_1$\\
$B_2+C_1$\\

$B_3+D_2$\\
$B_4+E_2$\\
$B_5+F_2$\\
$B_6+G_1$\\
$C_2+G_2$\\
$D_3+G_3$\\
$E_3+G_4$\\
$F_3+G_5$\\
$A_6+C_3+D_4$\\
$A_7+C_4+E_4$\\
$A_8+C_5+F_4$\\
$A_{9}+C_{6}+G_{6}$\\
$A_{10}+D_5+E_5$\\
$A_{11}+D_6+F_5$\\
$A_{12}+E_{6}+F_{6}$\\
$B_{7}+C_{7}+D_{7}$\\
$B_{8}+C_{8}+E_{7}$\\
$B_{9}+C_{9}+F_{7}$\\
$B_{10}+D_{8}+E_{8}$\\
$B_{11}+D_{9}+F_{8}$\\
$B_{12}+E_{9}+F_{9}$\\
$C_{10}+D_{10}+E_{10}$\\
$C_{11}+D_{11}+F_{10}$\\
$C_{12}+D_{12}+G_{7}$\\
$C_{13}+E_{11}+F_{11}$\\
$C_{14}+E_{12}+G_{8}$\\
$C_{15}+F_{12}+G_{9}$\\
$D_{13}+E_{13}+F_{13}$\\
$D_{14}+E_{14}+G_{10}$\\
$D_{15}+F_{14}+G_{11}$\\
$E_{15}+F_{15}+G_{12}$\\
$A_{13}+B_{13}+C_{16}+G_{13}$\\
$A_{14}+B_{14}+D_{16}+G_{14}$\\
$A_{15}+B_{15}+E_{16}+G_{15}$\\

$A_{16}+B_{16}+F_{16}+G_{16}$\\
$C_{17}+D_{17}+E_{17}+F_{17}$\\
$A_{17}+B_{17}+C_{18}+D_{18}+E_{18}$\\
$A_{18}+B_{18}+C_{19}+D_{19}+F_{18}$\\
$A_{19}+B_{19}+C_{20}+D_{20}+G_{17}$\\
$A_{20}+B_{20}+C_{21}+E_{19}+F_{19}$\\
$A_{21}+B_{21}+C_{22}+E_{20}+G_{18}$\\

$A_{22}+B_{22}+C_{23}+F_{20}+G_{19}$\\
$A_{23}+B_{23}+D_{21}+E_{21}+F_{21}$\\
$A_{24}+B_{24}+D_{22}+E_{22}+G_{20}$\\
$A_{25}+B_{25}+D_{23}+F_{22}+G_{21}$\\
$A_{26}+B_{26}+E_{23}+F_{23}+G_{22}$\\
$A_{27}+C_{24}+D_{24}+E_{24}+G_{23}$\\
$A_{28}+C_{25}+D_{25}+F_{24}+G_{24}$\\
$A_{29}+C_{26}+E_{25}+F_{25}+G_{25}$\\
$A_{30}+D_{26}+E_{26}+F_{26}+G_{26}$\\
$B_{27}+C_{27}+D_{27}+E_{27}+G_{27}$\\
$B_{28}+C_{28}+D_{28}+F_{27}+G_{28}$\\
$B_{29}+C_{29}+E_{28}+F_{28}+G_{29}$\\
$B_{30}+D_{29}+E_{29}+F_{29}+G_{30}$\\
$A_{31}+B_{31}+C_{30}+D_{30}+E_{30}+F_{30}$\\
$A_{32}+C_{31}+D_{31}+E_{31}+F_{31}+G_{31}$\\
$B_{32}+C_{32}+D_{32}+E_{32}+F_{32}+G_{32}$\\
 \hline
 \caption{Query to $N_{1}$ with index on each message}
 \label{tab:18}

\end{longtable}
\end{center}
\item 
The unknown byproduct combinations and the known byproduct bits of byproduct combinations from database $N_1$ are shown below. As per Proposition \ref{prop:symm_known_unknown}, observe that their structures are symmetric.
\begin{center}
\begin{longtable}{ |c|c| } 
\hline
UNKNOWN BYPRODUCT & KNOWN BYPRODUCT\\
\hline
$D_1$&$	D_{2}$\\
$D_4$&$	D_{7}$\\
$E_1$&$	E_{2}$\\
$E_4$&$	E_{7}$\\
$F_1$&$	F_{2}$\\
$F_4$&$	F_{7}$\\
$G_6$&$	G_{1}$\\
$G_{13}$&$	G_{2}$\\
$D_5+E_5$&$	D_{8}+E_{8}$\\
$D_{18}+E_{18}	$&$D_{10}+E_{10}$\\
$D_6+F_5	$&$D_{9}+F_{8}$\\
$D_{19}+F_{18}$&$D_{11}+F_{10}$\\
$D_{16}+G_{14}	$&$D_{3}+G_{3}$\\
$D_{20}+G_{17}$&$	D_{12}+G_{7}$\\
$E_6+F_6	$&$E_{9}+F_{9}$\\
$E_{19}+F_{19}$&$	E_{11}+F_{11}$\\
$E_{16}+G_{15}$&$	E_{3}+G_{4}$\\

$E_{20}+G_{18}	$&$E_{12}+G_{8}$\\
$F_{16}+G_{16}	$&$F_{3}+G_{5}$\\
$F_{20}+G_{19}$&$	F_{12}+G_{9}$\\
$D_{21}+E_{21}+F_{21}$&$	D_{13}+E_{13}+F_{13}$\\
$D_{30}+E_{30}+F_{30}$&$	D_{17}+E_{17}+F_{17}$\\
$D_{22}+E_{22}+G_{20}$&$	D_{14}+E_{14}+G_{10}$\\
$D_{24}+E_{24}+G_{23}$&$	D_{27}+E_{27}+G_{27}$\\
$D_{23}+F_{22}+G_{21}$&$	D_{15}+F_{14}+G_{11}$\\
$D_{25}+F_{24}+G_{24}$&$	D_{28}+F_{27}+G_{28}$\\
$E_{23}+F_{23}+G_{22}$&$	E_{15}+F_{15}+G_{12}$\\
$E_{25}+F_{25}+G_{25}$&$	E_{28}+F_{28}+G_{29}$\\
$D_{26}+E_{26}+F_{26}+G_{26}	$&$D_{29}+E_{29}+F_{29}+G_{30}$\\
$D_{31}+E_{31}+F_{31}+G_{31}$&$ D_{32}+E_{32}+F_{32}+G_{32}$\\
 \hline
 \caption{Known and unknown byproduct combinations of the query to $N_{1}$}
 \label{tab:19}

\end{longtable}
\end{center}
To generate the query for $N_2$, the structure of $N_1$ is copied without the indexes. The indexes of $A$ in $N_{1}$ are from 1 to 32. For the indexes of $A$ in ${N}_{2}$, use the indexes from 33 to  64. The indexes of $B,C$ are maintained as they are. Based on the one-to-one mapping in Table \ref{tab:19}, unknown and known byproduct combinations are swapped. Finally, the query for $N_{2}$ is as shown in Table \ref{tab:20}. Overall, 64 bits of $A$ are retrieved from both the databases by downloading a total of 126 bits. Thus, the rate of the code is $\frac{32}{63}$. Note that this rate is in between that of \cite{zchen} and \cite{sunjafar}  which had a rate of $\frac{64}{127}$ and $\frac{16}{31}$ respectively for the same parameters.

\begin{center}
\begin{longtable}{ |c|c| } 
\hline
QUERY TO $N_1$&QUERY TO $N_2$\\
\hline
$A_1$&	$A_{33}$\\		
$A_2+B_1$&	$A_{34}+B_1$\\		
$A_3+D_1$&	$A_{35}+D_2$	\\	
$A_4+E_1$&	$A_{36}+E_2$\\		
$A_5+F_1$&	$A_{37}+F_2$\\		
$B_2+C_1$&	$B_2+C_1$\\		
$B_3+D_2$&	$B_3+D_1$\\		
$B_4+E_2$&	$B_4+E_1$\\		
$B_5+F_2$&	$B_5+F_1$\\		
$B_6+G_1$&	$B_6+G_6$\\		
$C_2+G_2$&	$C_2+G_{13}$\\		
$D_3+G_3$&	$D_{16}+G_{14}$\\	
$E_3+G_4$&	$E_{16}+G_{15}$	\\	
$F_3+G_5$&	$F_{16}+G_{16}$\\		
$A_6+C_3+D_4$&	$A_{38}+C_3+D_7$	\\	
$A_7+C_4+E_4$&	$A_{39}+C_4+E_7$\\		
$A_8+C_5+F_4$&	$A_{40}+C_5+F_7$\\		
$A_{9}+C_{6}+G_{6}$&	$A_{41}+C_{6}+G_{1}$\\		
$A_{10}+D_5+E_5$&	$A_{42}+D_8+E_8$	\\	
$A_{11}+D_6+F_5$&	$A_{43}+D_9+F_8$\\		

$A_{12}+E_{6}+F_{6}$&	$A_{44}+E_{9}+F_{9}$\\		
$B_{7}+C_{7}+D_{7}$&	$B_{7}+C_{7}+D_{4}$	\\	
$B_{8}+C_{8}+E_{7}$&	$B_{8}+C_{8}+E_{4}$	\\	
$B_{9}+C_{9}+F_{7}$&	$B_{9}+C_{9}+F_{4}$	\\	
$B_{10}+D_{8}+E_{8}$&	$B_{10}+D_{5}+E_{5}$\\		
$B_{11}+D_{9}+F_{8}$&	$B_{11}+D_{6}+F_{5}$\\	
$B_{12}+E_{9}+F_{9}$&	$B_{12}+E_{6}+F_{6}$\\		
$C_{10}+D_{10}+E_{10}$&	$C_{10}+D_{18}+E_{18}$	\\	
$C_{11}+D_{11}+F_{10}$&	$C_{11}+D_{19}+F_{18}$	\\	
$C_{12}+D_{12}+G_{7}$&	$C_{12}+D_{20}+G_{17}$	\\	
$C_{13}+E_{11}+F_{11}$&	$C_{13}+E_{19}+F_{19}$	\\	
$C_{14}+E_{12}+G_{8}$&	$C_{14}+E_{20}+G_{18}$\\		
$C_{15}+F_{12}+G_{9}$&	$C_{15}+F_{20}+G_{19}$\\		
$D_{13}+E_{13}+F_{13}$&	$D_{21}+E_{21}+F_{21}$	\\	
$D_{14}+E_{14}+G_{10}$&	$D_{22}+E_{22}+G_{20}$	\\	
$D_{15}+F_{14}+G_{11}$&	$D_{23}+F_{22}+G_{21}$\\		
$E_{15}+F_{15}+G_{12}$&	$E_{23}+F_{23}+G_{22}$	\\	
$A_{13}+B_{13}+C_{16}+G_{13}$&	$A_{45}+B_{13}+C_{16}+G_{2}$	\\	
$A_{14}+B_{14}+D_{16}+G_{14}$&	$A_{46}+B_{14}+D_{3}+G_{3}$\\		
$A_{15}+B_{15}+E_{16}+G_{15}$&	$A_{47}+B_{15}+E_{3}+G_{4}$\\		
$A_{16}+B_{16}+F_{16}+G_{16}$&	$A_{48}+B_{16}+F_{3}+G_{5}$\\		
$C_{17}+D_{17}+E_{17}+F_{17}$&	$C_{17}+D_{30}+E_{30}+F_{30}$\\		
$A_{17}+B_{17}+C_{18}+D_{18}+E_{18}$&	$A_{49}+B_{17}+C_{18}+D_{10}+E_{10}$\\		
$A_{18}+B_{18}+C_{19}+D_{19}+F_{18}$&	$A_{50}+B_{18}+C_{19}+D_{11}+F_{10}$\\		
$A_{19}+B_{19}+C_{20}+D_{20}+G_{17}$&	$A_{51}+B_{19}+C_{20}+D_{12}+G_{7}$\\		
$A_{20}+B_{20}+C_{21}+E_{19}+F_{19}$&	$A_{52}+B_{20}+C_{21}+E_{11}+F_{11}$\\		
$A_{21}+B_{21}+C_{22}+E_{20}+G_{18}$&	$A_{53}+B_{21}+C_{22}+E_{12}+G_{8}$\\		
$A_{22}+B_{22}+C_{23}+F_{20}+G_{19}$&	$A_{54}+B_{22}+C_{23}+F_{12}+G_{9}$\\		
$A_{23}+B_{23}+D_{21}+E_{21}+F_{21}$&	$A_{55}+B_{23}+D_{13}+E_{13}+F_{13}$\\		
$A_{24}+B_{24}+D_{22}+E_{22}+G_{20}$&	$A_{56}+B_{24}+D_{14}+E_{14}+G_{10}$\\		
$A_{25}+B_{25}+D_{23}+F_{22}+G_{21}$&	$A_{57}+B_{25}+D_{15}+F_{14}+G_{11}$\\		
$A_{26}+B_{26}+E_{23}+F_{23}+G_{22}$&	$A_{58}+B_{26}+E_{15}+F_{15}+G_{12}$\\		
$A_{27}+C_{24}+D_{24}+E_{24}+G_{23}$&	$A_{59}+C_{24}+D_{27}+E_{27}+G_{27}$\\		
			
$A_{28}+C_{25}+D_{25}+F_{24}+G_{24}$&	$A_{60}+C_{25}+D_{28}+F_{27}+G_{28}$\\
$A_{29}+C_{26}+E_{25}+F_{25}+G_{25}$&	$A_{61}+C_{26}+E_{28}+F_{28}+G_{29}$\\		
$A_{30}+D_{26}+E_{26}+F_{26}+G_{26}$&	$A_{62}+D_{29}+E_{29}+F_{29}+G_{30}$\\		
$B_{27}+C_{27}+D_{27}+E_{27}+G_{27}$&	$B_{27}+C_{27}+D_{24}+E_{24}+G_{23}$\\		
$B_{28}+C_{28}+D_{28}+F_{27}+G_{28}$&	$B_{28}+C_{28}+D_{25}+F_{24}+G_{24}$\\		
$B_{29}+C_{29}+E_{28}+F_{28}+G_{29}$&	$B_{29}+C_{29}+E_{25}+F_{25}+G_{25}$\\		
$B_{30}+D_{26}+E_{26}+F_{26}+G_{26}$&	$B_{30}+D_{29}+E_{29}+F_{29}+G_{30}$\\		
$A_{31}+B_{31}+C_{30}+D_{30}+E_{30}+F_{30}$&	$A_{63}+B_{31}+C_{30}+D_{17}+E_{17}+F_{17}$	\\	

$A_{32}+C_{31}+D_{31}+E_{31}+F_{31}+G_{31}$&	$A_{64}+C_{31}+D_{32}+E_{32}+F_{32}+G_{32}$	\\	
$B_{32}+C_{32}+D_{32}+E_{32}+F_{32}+G_{32}$&	$B_{32}+C_{32}+D_{31}+E_{31}+F_{31}+G_{31}$	\\	
\hline
\caption{Queries to $N_{1}$ and $N_{2}$ for $K = 7$}
\label{tab:20}

\end{longtable}
\end{center}
\end{enumerate}

\vspace{-1cm}

\section{Proof for Joint Privacy of XOR-based PIR-PSI Codes}
\label{sec:4}

In this section, we provide a proof to show that the proposed code construction satisfies the design criteria for XOR-based PIR-PSI codes as listed in Section \ref{subsec:design_criteria}. Our approach is to use the code construction that is designed for the case when $X_{1}$ is the demand and $X_{2}, X_{3}$ are the side information messages. In particular, for this code, we fix the query submitted to $N_{1}$, and then show by construction that a query to $N_{2}$ can be generated in such a way that any demand can be retrieved from the two databases by using any two messages as the side information. Considering the case of arbitrary values of $K$, constructions are provided in three parts, namely: 
\begin{itemize}
    \item By fixing $X_{K}$ as one of the side information messages, we show that a query to $N_{2}$ can be synthesized such that any two messages out of the remaining $K-1$ messages can play the role of the demand and the other side information.
    \item By fixing $X_{K}$ as the demand, we show that a query to $N_{2}$ can be synthesized such that any two messages out of the remaining $K-1$ messages can play the role of side information. 
    \item By fixing $X_{K}$ as one of the byproducts, we show that a query to $N_{2}$ can be synthesized such that any three messages out of the remaining $K-1$ messages can play the role of the demand and two side information.
\end{itemize}

We start with the construction when $X_{K}$ is a side information. Since the code construction is presented by adding $2^{K-2}$ copies of $X_K$ to $P(\{X_{1}, X_{2}, \ldots, X_{K-1}\})$\textbackslash $\{\phi\}$, the privacy proof for this case is directly dependent on the power set structure of $P(\{X_{1}, X_{2}, \ldots, X_{K-1}\})$\textbackslash $\{\phi\}$. Given the power set structure, for any $K-3$ byproducts bits, the known and the unknown byproduct combinations in the query to $N_{1}$ are symmetric. As a result, the query submitted to $N_{2}$ must follow the same structure as that of $N_{1}$ except that (i) the index values of the demand will change from $2^{K-2} + 1$ to $2^{K-1}$ instead of $1$ to $2^{K-2}$, (ii) the known and unknown byproduct combinations are swapped similar to the proposed construction, and (iii) the index values of all the side information can be retained as they were in $N_{1}$. The method used for constructing query to $N_{2}$ under this case is summarized in the first row of Table \ref{tab:general1}. 

When $X_{K}$ is the demand message, for any given combination of the side information, i.e., $X_{i}X_{j}$ such that $i \neq j$ and $i, j \in [K]$\textbackslash$\{K\}$, we first list the corresponding known and unknown byproduct combinations using the existing query at $N_{1}$. Note that these byproduct combinations will involve the bits of messages other than that of $X_{K}, X_{i}, X_{j}$. Subsequently, we pick the query submitted to $N_{1}$, and then apply a suitable transformation between $\{X_{1}, X_{2}, X_{3}\}$ and $\{X_{1}, X_{2}, \ldots, X_{K-1}\}$ such that the known and the unknown byproduct combinations of the modified query are symmetric. Finally, if the known (or the unknown) byproduct combination in the query to $N_{1}$ is available as an unknown (or known) byproduct in the modified query, we swap their indexes, otherwise, we perform suitable modifications to the modified query such that the demand $X_{K}$ can be retrieved. Given that the proposed code construction is applicable when $X_{1}$ is the demand and $X_{2}, X_{3}$ are the side information, the same set of transformations is applicable on the query to $N_{1}$ for the side information messages within the following classes, namely: (i) the side information is one of $\{X_{1}X_{2}, X_{2}X_{3}\}$, (ii) the side information is one of $\{X_{1}X_{i} ~|~ 2 \leq i \leq K-1, i \neq 3\}$, (iii) the side information is one of $\{X_{2}X_{i} ~|~ 3 \leq i \leq K-1\}$, (iv) the side information is one of $\{X_{3}X_{i} ~|~ 4 \leq i \leq K-1\}$, and (v) the side information is one of $\{X_{i}X_{j} ~|~ 4 \leq i, j \leq K-1, i \neq j \}$. For each of these cases, the set of transformations that must be carried on the query submitted to $N_{1}$ is presented in the third column of Table \ref{tab:general1}, whereas the set of manipulations that must be applied after the transformation is listed in the fourth column of Table \ref{tab:general1}. It is important to note that after applying the transformations in the third column, a subset of the unknown and known byproduct combinations of the query to $N_{1}$ would be symmetric with that of the transformed query. On those subsets, the known and unknown byproduct combinations must be swapped similar to the case when $X_{K}$ plays the role of a side information. 

Finally, when $X_{K}$ is a byproduct, for a given demand and side information messages, we propose a sequence of transformations between the messages in the query to $N_{1}$ to ensure that the known and the unknown byproduct combinations match as much as that with that of the query to $N_{1}$. For the cases when the known and unknown byproduct combinations do not match, we propose modifications on the transformed query so as to retrieve all the bits of the demand. Table \ref{tab:general2} lists the instructions to obtain the query at $N_2$ for the case when $X_K$ is one of the byproducts.

In both Table \ref{tab:general1} and Table \ref{tab:general2}, the operator $\rightleftharpoons$ represents swapping the elements on either side of the $\rightleftharpoons$ operator, i.e., swap all the position of element in left hand side of $\rightleftharpoons$ with the one in its right hand side in the given query. For example, $A \rightleftharpoons B$ implies that swap all positions of $A$ and $B$ in the given query. If the $\rightleftharpoons$ operator has two juxtaposed  elements in either side, swap first and second element of the left side with first and second element of the right side, respectively. Similarly, the operator $\Rightarrow$ represents the replacement of the codeword in the left of the $\Rightarrow$ operator with the one in the right. For example, $A+B \Rightarrow C+D$ implies that, replace the codeword $A+B$ with $C+D$ in the given query. In summary, the first row of Table \ref{tab:general1}, shows that when $X_K$ is a side information, any demand can be retrieved with any side information pair in $N_2$ while keeping the query at $N_1$ unaltered and making necessary changes in $N_2$ without altering the structure of code in $N_1$. Similarly, Table \ref{tab:general1} and Table \ref{tab:general1} show that the same is possible when $X_K$ is the demand and $X_K$ is one of the byproduct, respectively. This shows that the query obtained from the code construction algorithm in Section \ref{sec:3} can be used to retrieve any demand with any pair of side information in $N_2$ while keeping the query at $N_1$ unaltered.

\newpage
\begin{table}[H]
    \centering

   \begin{tabular}{|c|c|c|c|}
    \hline
       No.&Case&Transformation&Manipulations \\
         
         \hline
 1&$X_K$ as SI&None&Swap the known and unknown \\
 &&& byproducts to obtain query at $N_2$.\\
 \hline
 2&$X_K$ as demand, &None&Swap the known and unknown \\
  &$X_1X_3$ or $X_2X_3$ are SI&& byproducts to obtain query at $N_2$.\\
       \hline
      
  3&$X_K$ as demand &Pick query at &Swap the singleton bit $X_1$ \\
 &$X_1X_i$ are SI &$N_1$ and apply&and the bit $X_3$ from\\
 &$2 \leq i \leq K-1$, $i\neq3$&$X_i \rightleftharpoons X_3$& the two-tuple sum $X_3+X_K$\\
 \hline
  4&$X_K$ as demand &Pick query at&  $X_3+X_K$ $\Rightarrow$ $X_2+X_K$\\
 &$X_2X_i$ are SI &$N_1$ and apply &$X_2$ $\Rightarrow$ $X_K$\\
 &$4 \leq i \leq K-1$ &$X_1X_3$ $\rightleftharpoons$ $X_2X_i$& $X_2+X_1+X_i+X_K$ $\Rightarrow$ $X_2+X_1+X_i+X_3$\\
 \hline
5&$X_K$ as demand &Pick query at&If $i=4$ and $K \leq 7$, perform the following: \\
 &$X_3X_i$ are SI &$N_1$ and apply  & $X_4$ $\Rightarrow$ $X_K$, $X_2+X_4$ $\Rightarrow$ $X_3+X_4$, $X_2+X_3$ $\Rightarrow$ $X_5+X_3$, \\											
&$4 \leq i \leq K-1$ &$X_1$ $\Rightarrow$ $X_4$&    $X_2+X_6$ $\Rightarrow$ $X_5+X_6$, $X_2+X_K$ $\Rightarrow$ $X_4+X_K$,$X_5+X_K$ $\Rightarrow$ \\											
&&&  $X_5+X_1$, $X_6+X_K$ $\Rightarrow$ $X_6+X_3$, $X_4+X_1+X_5$ $\Rightarrow$ \\											
&&& $X_4+X_1+X_2$, $X_4+X_5+X_6$ $\Rightarrow$ $X_4+X_K+X_6$, \\											
&&&$X_3+X_1+X_5$ $\Rightarrow$ $X_2+X_K+X_5$, $X_3+X_1+X_6$ $\Rightarrow$ \\											
&&&  $X_2+X_K+X_4$, $X_3+X_1+X_K$ $\Rightarrow$ $X_2+X_K+X_3$,  \\											
&&&$X_3+X_5+X_6$ $\Rightarrow$  $X_3+X_2+X_4$, $X_3+X_6+X_K$ $\Rightarrow$   \\											
&&& $X_2+X_6+X_K$, $X_4+X_2+X_3+X_K$ $\Rightarrow$ $X_1+X_5+X_3+X_K$,   \\											
&&& $X_4+X_2+X_1+X_K$ $\Rightarrow$ $X_1+X_2+X_6+X_4$,    \\											
&&&$X_4+X_2+X_5+X_K$ $\Rightarrow$ $X_4+X_2+X_5+X_6$,     \\											
&&&$X_4+X_2+X_6+X_K$  $\Rightarrow$ $X_4+X_1+X_6+X_K$,     \\											
&&&$X_4+X_2+X_1+X_5+X_6$ $\Rightarrow$ $X_4+X_3+X_1+X_5+X_6$,    \\											
&&&$X_4+X_1+X_5+X_6+X_K$ $\Rightarrow$ $X_3+X_1+X_5+X_6+X_K$, \\											
&&&$X_4+X_3+X_1+X_5+X_6+X_K$  $\Rightarrow$   \\											
&&&$X_4+X_2+X_1+X_5+X_6+X_K$. For $i \neq 4$ and $K \leq 7$ \\											

&&& apply $X_4 \rightleftharpoons X_i$ to the query obtained after transformation \\
&&& in column 3 and perform the manipulations mentioned above.\\
&&& For $K>7$, rearrange the query in column 3 such a way that\\
&&& the first $2^{K-2}-1$ codewords do not contain $X_t$ where\\
&&&$t=max(\{4,5,\ldots,K-1\}),t\neq i$. Perform manipulations on\\
&&& these codewords by the steps proposed under the same  \\
&&&  scenario, however by treating  the message set as \\
&&& $\{X_1,X_2,\ldots, X_K\}$ \textbackslash $\{X_t\}$. For the remaining $2^{K-2}$ codewords  \\
&&& perform the following: Perform manipulations on all codewords \\
&&& of the form $X_{t}+Q+X_K$, where \\
&&& $Q \in P(\{X_1,X_2,\ldots,X_{K-1}\}$ \textbackslash $\{
X_t\})$ \textbackslash $\{\phi\}$ similar to the steps  \\
&&& proposed under the same scenario of $K-1$ messages,   \\
&&&i.e. perform the same manipulations done to the codewords of    \\
&&&the form $X_{t-1}+Q'+X_{K-1}$, where \\
&&& $Q' \in P(\{X_1,X_2,\ldots,X_{K-2}\}$ \textbackslash $\{
X_{t-1}\})$ \textbackslash $\{\phi\}$ when number of\\
&&&messages was $K-1$. Similarly the manipulations performed \\
&&& for the codewords of the form  $X_{t-1}+Q'$ when number of   \\
&&&messages was $K-1$ should be repeated for all codewords of   \\
&&& the form $X_{t}+Q$. If $K=8$, perform the following (else skip):  \\  
&&& $X_1+X_m+X_n+X_o+U \rightleftharpoons$ $X_m+X_n+X_o+U$  and  \\
&&&$X_2+X_m+X_n+X_o+U$ $\rightleftharpoons X_1+X_2+X_m+X_n+X_o+U $,   \\
&&&where $U \in $ $ P(\{X_3,X_i\})$, $4 \leq m,n,o \leq K-1, m \neq n \neq o \neq i$. \\



  \hline
 6&$X_K$ as demand &Pick query at & $X_p+X_q+X_1$ $\Rightarrow$ $X_q+X_3+X_1$ \\
 &$X_pX_q$ are SI &$N_2$ obtained  & $X_1+X_t+X_3+X_K$ $\Rightarrow$ $X_1+X_t+X_p+X_K$  \\
 &$4 \leq p,q \leq K-1$&from case 5& where $X_t$ is some byproduct \\
 & $p \neq q$& $X_3X_i \rightleftharpoons X_pX_q$&other than $X_1$ and $X_3$ .\\
 
  \hline
        
  \hline
    \end{tabular}
    \caption{Instructions to obtain query at $N_2$ when $X_K$ is either SI or demand}
    \label{tab:general1}
\end{table}
\newpage
\begin{table}[H]
    \centering

   \begin{tabular}{|c|c|c|c|}
    \hline
       No.&Case&Transformation&Manipulations \\
         
         \hline

 9&$X_1$ as demand, $X_2X_i$ &None&Swap the known and unknown \\
  &  are SI, $3 \leq i \leq K-1$&& byproducts to obtain query at $N_2$.\\
       \hline
       10&$X_1$ as demand, $X_pX_q$ &Pick query at&None \\
  &  are SI, $3 \leq p,q \leq K-1$&$N_2$ obtained & \\
  &$p \neq q$&from case 5&\\
  &&$X_3X_i \rightleftharpoons X_pX_q$&\\
  &&$X_1 \rightleftharpoons X_K$&\\
  \hline
11&$X_2$ as demand, $X_1X_i$ &None&Swap the known and unknown \\
  &  are SI, $3 \leq i \leq K-1$&& byproducts to obtain query at $N_2$.\\
        \hline
        
         12&$X_2$ as demand, $X_pX_q$ &Pick query at&None \\
  &  are SI, $3 \leq p,q \leq K-1$&$N_2$ obtained & \\
  &$p \neq q$&from case 5&\\
   &&$X_3X_i \rightleftharpoons X_pX_q$&\\
  &&$X_2 \rightleftharpoons X_K$&\\
  \hline
    13&$X_3$ as demand &Pick query at& None \\
 &$X_1X_i$ are SI &$N_2$ obtained& \\
 &$2 \leq i \leq K-1$, $i\neq3$&from Case 3 & \\
 &&$X_3 \rightleftharpoons X_K$&\\
 \hline
 14&$X_3$ as demand &Pick query at& None \\
 &$X_2X_i$ are SI &$N_2$ obtained& \\
 &$4 \leq i \leq K-1$, $i\neq3$&from Case 4 & \\
 &&$X_3 \rightleftharpoons X_K$&\\
 \hline
  15&$X_3$ as demand &Pick query at& $X_i+X_j \Rightarrow$ $X_K+X_j$, $X_K+X_1 \Rightarrow$ $X_K+X_i$, $X_K+X_2 \Rightarrow$ \\
 &$X_iX_j$ are SI &$N_2$ obtained& $X_3+X_2$, $X_K+X_3+X_2 \Rightarrow$ $X_j+X_3+X_2$,\\
 & $4 \leq i,j \leq K-1$, $i\neq j$ &from Case 10 &$X_2+X_K+X_j+X_3 \Rightarrow X_1+X_4+X_i+X_3$,   \\
 && $X_2 \rightleftharpoons X_3$& $X_j+X_1+X_4+X_3  \Rightarrow X_j+X_1+X_2+X_K$\\
 &&&$X_j+X_i+X_1+X_2+X_3 \Rightarrow X_1+X_K+X_2+X_4+X_j$\\
  &&&$X_j+X_K+X_i+X_1+X_2+X_4 \Rightarrow $\\
  &&&$X_1+X_2+X_K+X_3+X_i+X_j$\\
 \hline
 
 16&$X_\delta$ as demand, $X_iX_j$ &None&Swap the known and unknown \\
  &  are SI, $4 \leq \delta \leq K-1$&& byproducts to obtain query at $N_2$.\\
  &$1 \leq i \leq 2, 2 \leq j \leq K-1, $&&\\
  &$i \neq j,  i,j \neq \delta$&&\\
       \hline
 17&$X_\delta$ as demand, $X_pX_q$ &Pick query at&None \\
  &  are SI, $4 \leq \delta \leq K-1,$&$N_2$ obtained & \\
  &$3 \leq p,q \leq K-1$, $p \neq q$&from case 5&\\
   &$p,q \neq \delta$&$X_3X_i \rightleftharpoons X_pX_q$&\\
  &&$X_\delta \rightleftharpoons X_K$&\\
  
  \hline
   18&$X_\delta$ as demand, $X_pX_q$ &Pick query at&$X_1+X_\delta \Rightarrow X_2+X_\delta$, $X_2+X_\delta+X_p \Rightarrow X_1+X_K+X_\delta$, \\
  &  are SI, $4 \leq \delta \leq K-1,$&$N_2$ obtained & $X_1+X_K+X_q \Rightarrow X_1+X_q+X_p$\\
  &$4 \leq p,q \leq K-1$, $p \neq q$&from case 5&\\
   &$p,q \neq \delta$&$X_3X_i \rightleftharpoons X_pX_q$&\\
  &&$X_\delta \rightleftharpoons X_K$&\\
  
  \hline
    \end{tabular}
    \caption{Instructions to obtain query at $N_2$ when $X_K$ one of the byproducts}
    \label{tab:general2}
\end{table}

\section{Proof for Joint Privacy for the Code with $K = 7$}
\label{sec:privacy_proof_K_7}

In this section, we provide a proof for the code constructed in Section \ref{sec:3a}, i.e., for $K = 7$. In the process, the letter $G$ will play the role of $X_{K}$, whereas the letters $A, B, C, \ldots, F$ play the role of $X_{1}, X_{2}, \ldots, X_{K-1}$, respectively. The objective of this section is to show the execution of instructions presented in Table \ref{tab:general1} and Table \ref{tab:general2}.

\subsection{$G$ as a side information}
\label{sec:g1}

Recall that the queries for $K > 3$, $M=2$ is formed by adding $2^{K-2}$ copies of $X_K$ (in this case $G$ for the example in Section \ref{sec:3a}) to $P(\{X_{1}, X_{2}, \ldots, X_{K-1}\})$\textbackslash $\{\phi\}$ (in this case the power set structure $P(\{A, B, C, D, E, F\})$\textbackslash $\{\phi\}$). Since $G$ is the side information, it can be removed from downloaded bits, and therefore, the privacy proof of the code is directly dependent on the structure of $P(\{A, B, C, D, E, F\})$\textbackslash $\{\phi\}$. For any given demand and any side information from the remaining five messages, structure of $P(\{A, B, C, D, E, F\})$\textbackslash $\{\phi\}$ guarantees that the known byproduct combinations and the unknown byproduct combinations in the query to $N_{1}$ are symmetric. Hence, the query submitted to $N_{2}$ follows the same structure as that of $N_{1}$ except that 
\begin{itemize}
    \item The index values of the demand will change from 33 to 64 instead of 1 to 32. 
    \item The known and unknown byproduct combinations are swapped similar to the proposed construction.
    \item The index values of all the side information can be retained as they were in $N_{1}$.
\end{itemize}

\subsection{G as the demand}
\label{sec:g2}

When $G$ is the demand, the two side information messages can come from $\{A,B,\ldots,E,F\}$ in $K-1 \choose 2$ = $6 \choose 2$ = 15 ways. The 15 combinations are $\{AB,AC,AD,AE,AF,BC,BD,BE,BF,\allowbreak CD,CE,CF,DE,DF,EF\}$, implying that the two letters juxtaposed next to each other are the side information messages. These 15 combinations are grouped into five different types, namely: $\{AC,BC\}$, $\{AB,AD,AE,AF\}$, $\{BD,BE,BF\}$, $\{CD,CE,CF\}$ and $\{DE,DF,EF\}$. The reason for this classification is attributed to the fact that the query to $N_{1}$ was constructed assuming $A$ as the demand and $B, C$ as the side information, and therefore, when we have to design the new query to $N_{2}$, it should match with the existing query to $N_{1}$.\\

\subsubsection{One of $\{AC,BC\}$ as side information and $G$ as the demand}
\label{sec:g21}
    
When either $AC$ or $BC$ are the side information messages, the known and unknown byproduct combinations are symmetric. With the side information $AC$, the pattern of unknown and known byproduct combinations without the indexes are given in Table \ref{tab:21}. Therefore, the demand $G$ can be retrieved by swapping the indexes of the known and unknown byproduct combinations. Similar to the code construction, when submitting the query to $N_{2}$, the indexes on side information must be the same, whereas the indexes of the demand on $N_{2}$ must be new.
    \begin{center}
\begin{longtable}{ |c|c| }
\hline
UNKNOWN&KNOWN\\
\hline
B&B\\
B&B\\
D&D\\
D&D\\
E&E\\
E&E\\
F&F\\
F&F\\
B+D&B+D\\
B+D&B+D\\
B+E&B+E\\
B+E&B+E\\
B+F&B+F\\
B+F&B+F\\
D+E&D+E\\
D+E&D+E\\

D+F&D+F\\
D+F&D+F\\
E+F&E+F\\
E+F&E+F\\
B+D+E&B+D+E\\
B+D+E&B+D+E\\
B+D+F&B+D+F\\
B+D+F&B+D+F\\
B+E+F&B+E+F\\
B+E+F&B+E+F\\
D+E+F&D+E+F\\
D+E+F&D+E+F\\
B+D+E+F&B+D+E+F\\
B+D+E+F&B+D+E+F\\
\hline
\caption{Pattern of unknown-known byproducts for SI $AC$}
\label{tab:21}
\end{longtable}
\end{center}
In the case when $BC$ are the side information messages, the pattern of the known and the unknown byproduct combinations remains symmetric, and therefore, the query to $N_{2}$ is similar to that when $AC$ were the side information messages. The only exception is that the role of $A$ and $B$ gets swapped as $A$ becomes a byproduct and $B$ becomes a side information. Thus, all the bits of $G$ can be retrieved from both databases without altering the structure of query to $N_1$ when one of $\{AC,BC\}$ are the side information messages.


\subsubsection{One of $\{AB,AD,AE,AF\}$ as side information and $G$ as the demand}  
\label{sec:g22}

\noindent In the case when one of $\{AB,AD,AE,AF\}$ are the side information, the known and unknown byproduct combinations are almost symmetric with one exception that a singleton bit of byproduct $C$ goes to the unknown side from the known side. For instance, when $AB$ are the side information, the known and unknown byproduct combinations are as shown in the first two columns of Table \ref{tab:22}. Therefore, the query to $N_2$ should somehow accommodate this extra unknown bit without changing the structure. To construct a query to $N_{2}$, we use the query to $N_{1}$ and make appropriate modifications as discussed hereafter. For exposition, we take the case when $AB$ are the side information. We already know that the query to $N_{1}$ gives a symmetric known-unknown byproduct combinations for side information $AC$ and demand $G$. We pick the query submitted to $N_{1}$, and propose a simple transformation of $B$ $\rightleftharpoons$ $C$, i.e., swapping all positions of $B$ with $C$, to generate a new query. Due to the transformation, it is clear that this new query gives a symmetric known-unknown byproduct combinations when $G$ is the demand and $AB$  are the side information. This pattern of unknown and known byproduct combinations are shown in the third and the fourth columns of Table \ref{tab:22}.

\begin{small}
\begin{center}
\begin{longtable}{ |c|c|c|c| }
\hline
UNKNOWN&KNOWN&UNKNOWN &KNOWN \\
&&FOR $B$ $\rightleftharpoons$ $C$&FOR $B$ $\rightleftharpoons$ $C$\\
\hline
C&&&			\\
C&&	C&C		\\
C&C&	C&C		\\
D&D&	D&D		\\
D&D&	D&D		\\
E&E&	E&E		\\
E&E&	E&E		\\
F&F&	F&F		\\
F&F&	F&F		\\
C+D&C+D&	C+D&C+D		\\
C+D&C+D&	C+D&C+D		\\
C+E&C+E&	C+E&C+E		\\
C+E&C+E&	C+E&C+E		\\
C+F&C+F&	C+F&C+F		\\
C+F&C+F&	C+F&C+F		\\
D+E&D+E&	D+E&D+E		\\
D+E&D+E&	D+E&D+E		\\
D+F&D+F&	D+F&D+F		\\
D+F&D+F&	D+F&D+F		\\
E+F&E+F&	E+F&E+F		\\
E+F&E+F&	E+F&E+F		\\
C+D+E&C+D+E&	C+D+E&C+D+E		\\
C+D+E&C+D+E&	C+D+E&C+D+E		\\
C+D+F&C+D+F&	C+D+F&C+D+F		\\
C+D+F&C+D+F&	C+D+F&C+D+F		\\
C+E+F&C+E+F&	C+E+F&C+E+F		\\
C+E+F&C+E+F&	C+E+F&C+E+F		\\
D+E+F&D+E+F&	D+E+F&D+E+F		\\
D+E+F&D+E+F&	D+E+F&D+E+F		\\
C+D+E+F&C+D+E+F&	C+D+E+F&C+D+E+F		\\
C+D+E+F&C+D+E+F&	C+D+E+F&C+D+E+F		\\
			
\hline
\caption{Pattern of unknown-known byproducts for SI $AB$ and after $B$ $\rightleftharpoons$ $C$}
\label{tab:22}
\end{longtable}
\end{center}
\end{small}

We now provide modifications on the new query after the transformation $B$ $\rightleftharpoons$ $C$ so that we can use as our query to $N_{2}$. To help this cause, the original query to $N_1$ for $K=7$ (without indexes) with numbering for each codeword is given in Table \ref{tab:23}.

\begin{center}
\begin{longtable}{ |c|c| }			
\hline			
\multicolumn{2}{|c|}{BIT NUMBER TABLE}\\
\hline
BIT NUMBER&CODE\\			
\hline			
1	&	A\\
2	&	A+B	\\
3	&	A+D	\\
4	&	A+E	\\
5	&	A+F	\\

6	&	B+C	\\
7	&	B+D	\\
8	&	B+E	\\
9	&	B+F	\\
10	&	B+G	\\
11	&	C+G	\\
12	&	D+G	\\
13	&	E+G	\\
14	&	F+G	\\
15	&	A+C+D	\\
16	&	A+C+E	\\
17	&	A+C+F	\\
18	&	A+C+G	\\
19	&	A+D+E	\\
20	&	A+D+F	\\
21	&	A+E+F	\\
22	&	B+C+D	\\
23	&	B+C+E	\\
24	&	B+C+F	\\
25	&	B+D+E	\\
26	&	B+D+F	\\
27	&	B+E+F	\\
28	&	C+D+E	\\
29	&	C+D+F	\\
30	&	C+D+G	\\
31	&	C+E+F	\\
32	&	C+E+G	\\
33	&	C+F+G	\\
34	&	D+E+F	\\
35	&	D+E+G	\\
36	&	D+F+G	\\
37	&	E+F+G	\\
38	&	A+B+C+G	\\
39	&	A+B+D+G	\\
40	&	A+B+E+G	\\
41	&	A+B+F+G	\\
42	&	C+D+E+F	\\
43	&	A+B+C+D+E	\\
44	&	A+B+C+D+F	\\
45	&	A+B+C+D+G	\\
46	&	A+B+C+E+F	\\

47	&	A+B+C+E+G	\\
48	&	A+B+C+F+G	\\
49	&	A+B+D+E+F	\\
50	&	A+B+D+E+G	\\
51	&	A+B+D+F+G	\\
52	&	A+B+E+F+G	\\
53	&	A+C+D+E+G	\\
54	&	A+C+D+F+G	\\
55	&	A+C+E+F+G	\\
56	&	A+D+E+F+G	\\
57	&	B+C+D+E+G	\\
58	&	B+C+D+F+G	\\
59	&	B+C+E+F+G	\\
60	&	B+D+E+F+G	\\
61	&	A+B+C+D+E+F	\\
62	&	A+C+D+E+F+G	\\
63	&	B+C+D+E+F+G	\\
\hline		
\caption{Bit number table maps a bit to a number for reference}
\label{tab:23}
\end{longtable}			
\end{center}			

Since one bit of $C$ moved to the unknown side from the known side in the existing query to $N_{1}$, we pick the new query (after the transformation) and then swap the singleton bit $A$ ( at bit number 1) and the bit $C$ from the two-tuple sum $C+G$ (bit number 10). This ensures that the extra unknown bit $C$ is retrieved in bit number 1 while still being able to retrieve the demand $G$ bit in bit number 10 since $A$ is a side information. This sequence of operations from taking a copy of the query to $N_{1}$, the transformation of $B$ $\rightleftharpoons$ $C$, and the final exchange in bit positions 1 and 10 are displayed in the last three columns of Table \ref{tab:24}. The rows of Table \ref{tab:24} indicated in red are the ones that get modified. Thus, the last column of this table forms the query to $N_{2}$ when $AB$ are the side information. Of course, swapping of the known-unknown byproducts indexes is required wherever they are symmetric. 

\begin{center}							
\begin{longtable}{ |c|c|c|c| }							
\hline							
BIT NUMBER&$N_1$&$B$ $\rightleftharpoons$ $C$&$N_2$\\						
\hline							
\textcolor{red}{1}	&	\textcolor{red}{A}	&	\textcolor{red}{A}	&	\textcolor{red}{C}\\
2	&	A+B	&	A+C	&	A+C	\\
3	&	A+D	&	A+D	&	A+D	\\
4	&	A+E	&	A+E	&	A+E	\\
5	&	A+F	&	A+F	&	A+F	\\
6	&	B+C	&	C+B	&	C+B	\\
7	&	B+D	&	C+D	&	C+D	\\
8	&	B+E	&	C+E	&	C+E	\\

9	&	B+F	&	C+F	&	C+F	\\
\textcolor{red}{10}	&	\textcolor{red}{B+G}	&	\textcolor{red}{C+G}	&	\textcolor{red}{A+G}	\\
11	&	C+G	&	B+G	&	B+G	\\
12	&	D+G	&	D+G	&	D+G	\\
13	&	E+G	&	E+G	&	E+G	\\
14	&	F+G	&	F+G	&	F+G	\\
15	&	A+C+D	&	A+B+D	&	A+B+D	\\
16	&	A+C+E	&	A+B+E	&	A+B+E	\\
17	&	A+C+F	&	A+B+F	&	A+B+F	\\
18	&	A+C+G	&	A+B+G	&	A+B+G	\\
19	&	A+D+E	&	A+D+E	&	A+D+E	\\
20	&	A+D+F	&	A+D+F	&	A+D+F	\\
21	&	A+E+F	&	A+E+F	&	A+E+F	\\
22	&	B+C+D	&	C+B+D	&	C+B+D	\\
23	&	B+C+E	&	C+B+E	&	C+B+E	\\
24	&	B+C+F	&	C+B+F	&	C+B+F	\\
25	&	B+D+E	&	C+D+E	&	C+D+E	\\
26	&	B+D+F	&	C+D+F	&	C+D+F	\\
27	&	B+E+F	&	C+E+F	&	C+E+F	\\
28	&	C+D+E	&	B+D+E	&	B+D+E	\\
29	&	C+D+F	&	B+D+F	&	B+D+F	\\
30	&	C+D+G	&	B+D+G	&	B+D+G	\\
31	&	C+E+F	&	B+E+F	&	B+E+F	\\
32	&	C+E+G	&	B+E+G	&	B+E+G	\\
33	&	C+F+G	&	B+F+G	&	B+F+G	\\
34	&	D+E+F	&	D+E+F	&	D+E+F	\\
35	&	D+E+G	&	D+E+G	&	D+E+G	\\
36	&	D+F+G	&	D+F+G	&	D+F+G	\\
37	&	E+F+G	&	E+F+G	&	E+F+G	\\
38	&	A+B+C+G	&	A+C+B+G	&	A+C+B+G	\\
39	&	A+B+D+G	&	A+C+D+G	&	A+C+D+G	\\
40	&	A+B+E+G	&	A+C+E+G	&	A+C+E+G	\\
41	&	A+B+F+G	&	A+C+F+G	&	A+C+F+G	\\
42	&	C+D+E+F	&	B+D+E+F	&	B+D+E+F	\\
43	&	A+B+C+D+E	&	A+C+B+D+E	&	A+C+B+D+E	\\
44	&	A+B+C+D+F	&	A+C+B+D+F	&	A+C+B+D+F	\\
45	&	A+B+C+D+G	&	A+C+B+D+G	&	A+C+B+D+G	\\
46	&	A+B+C+E+F	&	A+C+B+E+F	&	A+C+B+E+F	\\
47	&	A+B+C+E+G	&	A+C+B+E+G	&	A+C+B+E+G	\\
48	&	A+B+C+F+G	&	A+C+B+F+G	&	A+C+B+F+G	\\
49	&	A+B+D+E+F	&	A+C+D+E+F	&	A+C+D+E+F	\\

50	&	A+B+D+E+G	&	A+C+D+E+G	&	A+C+D+E+G	\\
51	&	A+B+D+F+G	&	A+C+D+F+G	&	A+C+D+F+G	\\
52	&	A+B+E+F+G	&	A+C+E+F+G	&	A+C+E+F+G	\\
53	&	A+C+D+E+G	&	A+B+D+E+G	&	A+B+D+E+G	\\
54	&	A+C+D+F+G	&	A+B+D+F+G	&	A+B+D+F+G	\\
55	&	A+C+E+F+G	&	A+B+E+F+G	&	A+B+E+F+G	\\
56	&	A+D+E+F+G	&	A+D+E+F+G	&	A+D+E+F+G	\\
57	&	B+C+D+E+G	&	C+B+D+E+G	&	C+B+D+E+G	\\
58	&	B+C+D+F+G	&	C+B+D+F+G	&	C+B+D+F+G	\\
59	&	B+C+E+F+G	&	C+B+E+F+G	&	C+B+E+F+G	\\
60	&	B+D+E+F+G	&	C+D+E+F+G	&	C+D+E+F+G	\\
61	&	A+B+C+D+E+F	&	A+C+B+D+E+F	&	A+C+B+D+E+F	\\
62	&	A+C+D+E+F+G	&	A+B+D+E+F+G	&	A+B+D+E+F+G	\\
63	&	B+C+D+E+F+G	&	C+B+D+E+F+G	&	C+B+D+E+F+G	\\
\hline			
\caption{Final query to $N_2$ given in the fourth column.}
\label{tab:24}
\end{longtable}							
\end{center}	
In general, when handling the other cases of $\{AD,AE,AF\}$ as the side information, the procedure for obtaining query for $N_2$ is similar to that when $AB$ is the side information. We pick the query submitted to $N_1$ and perform the transformation of $C$ with $D,E,F$ for the cases $AD,AE,AF$, respectively. The modifications made to the new query after the transformation is the same as that when $AB$ were the side information. Finally, the bit numbers would change according to the new position of codeword $C+G$.


\subsubsection{One of $\{BD,BE,BF\}$ as side information and $G$ as the demand}
\label{sec:g23}

\noindent In the case when one of $\{BD,BE,BF\}$ are the side information, the known and unknown byproduct combinations are almost symmetric with two exceptions that a two-tuple sum of byproduct combination, $A+C$ goes to the unknown side from the known side while a bit of byproduct $A$ goes to the known side from the unknown side. For instance, when $BD$ are the side information,  the known and unknown byproduct combinations are as shown in the first two columns of Table \ref{tab:25}. Therefore, the query to $N_2$ should somehow accommodate these deviations without changing the structure. To construct a query to $N_{2}$, we use the query to $N_{1}$ and make appropriate modifications as discussed hereafter. For exposition, we take the case when $BD$ are the side information. We already know that the query to $N_{1}$ gives a symmetric known-unknown byproduct combinations for side information $AC$ and demand $G$. We pick the query submitted to $N_{1}$, and perform a transformation of $A$ $\rightleftharpoons$ $B$ and $C$ $\rightleftharpoons$ $D$, to generate a new query. Due to the transformation, it is clear that this new query gives a symmetric known-unknown byproduct combinations when $G$ is the demand and $BD$  are the side information. This pattern of unknown and known byproduct combinations are shown in the third and the fourth columns of Table \ref{tab:25}.

\begin{small}
\begin{center}
\begin{longtable}{ |c|c|c|c| }
\hline
UNKNOWN&KNOWN&UNKNOWN &KNOWN \\
&&FOR $A$ $\rightleftharpoons$ $B$&FOR $A$ $\rightleftharpoons$ $B$\\
&&$C$ $\rightleftharpoons$ $D$&$C$ $\rightleftharpoons$ $D$\\
\hline
&A&	&	\\
&A&	A&A	\\
A&A&	A&A	\\
C&C&	C&C	\\
C&C&	C&C	\\
E&E&	E&E	\\
E&E&	E&E	\\
F&F&	F&F	\\
F&F&	F&F	\\
A+C&A+C&	&	\\
A+C&&	A+C&A+C	\\
A+C&&	A+C&A+C	\\
A+E&A+E&	A+E&A+E	\\
A+E&A+E&	A+E&A+E	\\
A+F&A+F&	A+F&A+F	\\
A+F&A+F&	A+F&A+F	\\
C+E&C+E&	C+E&C+E	\\
C+E&C+E&	C+E&C+E	\\
C+F&C+F&	C+F&C+F	\\
C+F&C+F&	C+F&C+F	\\
E+F&E+F&	E+F&E+F	\\

E+F&E+F&	E+F&E+F	\\
A+C+E&A+C+E&	A+C+E&A+C+E	\\
A+C+E&A+C+E&	A+C+E&A+C+E	\\
A+C+F&A+C+F&	A+C+F&A+C+F	\\
A+C+F&A+C+F&	A+C+F&A+C+F	\\
A+E+F&A+E+F&	A+E+F&A+E+F	\\
A+E+F&A+E+F&	A+E+F&A+E+F	\\
C+E+F&C+E+F&	C+E+F&C+E+F	\\
C+E+F&C+E+F&	C+E+F&C+E+F	\\
A+C+E+F&A+C+E+F&	A+C+E+F&A+C+E+F	\\
A+C+E+F&A+C+E+F&	A+C+E+F&A+C+E+F	\\
\hline
\caption{Pattern of unknown and known byproducts for SI $BD$ and after  $A$ $\rightleftharpoons$ $B$ and $C$ $\rightleftharpoons$ $D$.}
\label{tab:25}

\end{longtable}
\end{center}
\end{small}

Since a bit of $A+C$ is moved to the unknown side from the known side in the existing query to $N_{1}$, there is only one known bit of $A+C$ available to retrieve $G$. But the query after transformation uses 2 bits of $A+C$ (bit number 39 and 45) to retrieve 2 bits of $G$ as seen in the third column of the Table \ref{tab:26}. Since we have one extra bit of $A$ available, to get the extra bit of $A+C$, the bit $C+G$ in bit number 12 is replaced with bit $B+G$. $G$ is still retrievable since $B$ is side information. The bit of $C$ freed from bit number 12 along with extra bit of $A$ contributes to the second $A+C$ bit. Since one more $B$ is added to the query, the singleton bit of $B$ is replaced with demand bit $G$ as seen in bit number 1 in the table below. Since one extra demand bit is retrieved, the bit number 38 is changed from $B+A+D+G$ to $B+A+D+C$ so that the extra unknown bit of $A+C$ is also acquired. Since one bit of $C$ was removed in bit number 12, this change will bring back the uniform distribution of bits while keeping the structure of query at $N_2$ maintained as that of $N_1$. This sequence of operations from taking a copy of the query to $N_{1}$, the transformation of $A$ $\rightleftharpoons$ $B$ and $C$ $\rightleftharpoons$ $D$, and the final exchange in bit positions 1, 12, and 38 are displayed in the last three columns of Table \ref{tab:26}. Thus, the last column of this table forms the query to $N_{2}$ when $BD$ are the side information. Of course, swapping of the known-unknown byproducts indexes is required wherever they are symmetric. 

\begin{center}							
\begin{longtable}{ |c|c|c|c| }							
\hline							
BIT NUMBER&$N_1$&$A$ $\rightleftharpoons$ $B$, $C$ $\rightleftharpoons$ $D$&$N_2$\\						
\hline							
\textcolor{red}{1}	&	\textcolor{red}{A}	&	\textcolor{red}{B}	&	\textcolor{red}{G}	\\
2	&	A+B	&	B+A	&	B+A	\\
3	&	A+D	&	B+C	&	B+C	\\
4	&	A+E	&	B+E	&	B+E	\\
5	&	A+F	&	B+F	&	B+F	\\
6	&	B+C	&	A+D	&	A+D	\\
7	&	B+D	&	A+C	&	A+C	\\

8	&	B+E	&	A+E	&	A+E	\\
9	&	B+F	&	A+F	&	A+F	\\
10	&	B+G	&	A+G	&	A+G	\\
11	&	C+G	&	D+G	&	D+G	\\
\textcolor{red}{12}	&	\textcolor{red}{D+G}	&	\textcolor{red}{C+G}	&	\textcolor{red}{B+G}	\\
13	&	E+G	&	E+G	&	E+G	\\
14	&	F+G	&	F+G	&	F+G	\\
15	&	A+C+D	&	B+D+C	&	B+D+C	\\
16	&	A+C+E	&	B+D+E	&	B+D+E	\\
17	&	A+C+F	&	B+D+F	&	B+D+F	\\
18	&	A+C+G	&	B+D+G	&	B+D+G	\\
19	&	A+D+E	&	B+C+E	&	B+C+E	\\
20	&	A+D+F	&	B+C+F	&	B+C+F	\\
21	&	A+E+F	&	B+E+F	&	B+E+F	\\
22	&	B+C+D	&	A+D+C	&	A+D+C	\\
23	&	B+C+E	&	A+D+E	&	A+D+E	\\
24	&	B+C+F	&	A+D+F	&	A+D+F	\\
25	&	B+D+E	&	A+C+E	&	A+C+E	\\
26	&	B+D+F	&	A+C+F	&	A+C+F	\\
27	&	B+E+F	&	A+E+F	&	A+E+F	\\
28	&	C+D+E	&	D+C+E	&	D+C+E	\\
29	&	C+D+F	&	D+C+F	&	D+C+F	\\
30	&	C+D+G	&	D+C+G	&	D+C+G	\\
31	&	C+E+F	&	D+E+F	&	D+E+F	\\
32	&	C+E+G	&	D+E+G	&	D+E+G	\\
33	&	C+F+G	&	D+F+G	&	D+F+G	\\
34	&	D+E+F	&	C+E+F	&	C+E+F	\\
35	&	D+E+G	&	C+E+G	&	C+E+G	\\
36	&	D+F+G	&	C+F+G	&	C+F+G	\\

37	&	E+F+G	&	E+F+G	&	E+F+G	\\
\textcolor{red}{38}	&	\textcolor{red}{A+B+C+G}	&	\textcolor{red}{B+A+D+G}	&	\textcolor{red}{B+A+D+C}	\\
39	&	A+B+D+G	&	B+A+C+G	&	B+A+C+G	\\
40	&	A+B+E+G	&	B+A+E+G	&	B+A+E+G	\\
41	&	A+B+F+G	&	B+A+F+G	&	B+A+F+G	\\
42	&	C+D+E+F	&	D+C+E+F	&	D+C+E+F	\\
43	&	A+B+C+D+E	&	B+A+D+C+E	&	B+A+D+C+E	\\
44	&	A+B+C+D+F	&	B+A+D+C+F	&	B+A+D+C+F	\\
45	&	A+B+C+D+G	&	B+A+D+C+G	&	B+A+D+C+G	\\
46	&	A+B+C+E+F	&	B+A+D+E+F	&	B+A+D+E+F	\\
47	&	A+B+C+E+G	&	B+A+D+E+G	&	B+A+D+E+G	\\
48	&	A+B+C+F+G	&	B+A+D+F+G	&	B+A+D+F+G	\\
49	&	A+B+D+E+F	&	B+A+C+E+F	&	B+A+C+E+F	\\
50	&	A+B+D+E+G	&	B+A+C+E+G	&	B+A+C+E+G	\\
51	&	A+B+D+F+G	&	B+A+C+F+G	&	B+A+C+F+G	\\
52	&	A+B+E+F+G	&	B+A+E+F+G	&	B+A+E+F+G	\\
53	&	A+C+D+E+G	&	B+D+C+E+G	&	B+D+C+E+G	\\
54	&	A+C+D+F+G	&	B+D+C+F+G	&	B+D+C+F+G	\\
55	&	A+C+E+F+G	&	B+D+E+F+G	&	B+D+E+F+G	\\
56	&	A+D+E+F+G	&	B+C+E+F+G	&	B+C+E+F+G	\\
57	&	B+C+D+E+G	&	A+D+C+E+G	&	A+D+C+E+G	\\
58	&	B+C+D+F+G	&	A+D+C+F+G	&	A+D+C+F+G	\\
59	&	B+C+E+F+G	&	A+D+E+F+G	&	A+D+E+F+G	\\
60	&	B+D+E+F+G	&	A+C+E+F+G	&	A+C+E+F+G	\\
61	&	A+B+C+D+E+F	&	B+A+D+C+E+F	&	B+A+D+C+E+F	\\
62	&	A+C+D+E+F+G	&	B+D+C+E+F+G	&	B+D+C+E+F+G	\\
63	&	B+C+D+E+F+G	&	A+D+C+E+F+G	&	A+D+C+E+F+G	\\
\hline			
\caption{Final query to $N_2$ given in the fourth column.}
\label{tab:26}
\end{longtable}							
\end{center}	

In general, when handling the other cases of $\{BE,BF\}$ as the side information, the procedure for obtaining query for $N_2$ is similar to that when $BD$ were the side information. We pick the query submitted to $N_1$ and perform the transformation between $A$ with $B$, and then between $C$ and $E$ (or $F$), for $BE$ (or $BF$) as the side information. The modifications made to the new query after the transformation when $BD$ were the side information can be reproduced to retrieve $G$ from $N_2$.


\subsubsection{One of $\{CD,CE,CF\}$ as side information and $G$ as the demand}
\label{sec:g24}

In the case when one of $\{CD,CE,CF\}$ are the side information, there is a drastic change in the known and unknown byproduct combinations compared to the previous cases when of $G$ is the demand. For instance, when $CD$ are the side information the singleton bits of byproducts $A$ and $B$ are unknown only one time whereas they are known for three times. The two-tuple sum block combinations of byproducts except $A+B$ are unknown only one time but are known three times whereas $A+B$ is unknown three times but known only one time. All the three-tuple sum blocks combinations of byproducts are unknown three times while they are known only once. The four-tuple sum block combinations of byproducts are known three times while they are unknown only once. The known and unknown byproduct combinations are as shown in Table \ref{tab:27}. 

\begin{center}
\begin{longtable}{ |c|c| }
\hline
UNKNOWN&KNOWN\\
\hline
A	&	A	\\
B	&	A	\\
E	&	A	\\
E	&	B	\\
E	&	B	\\
F	&	B	\\
F	&	E	\\

F	&	F	\\
A+B	&	A+B	\\
A+B	&	A+E	\\
A+B	&	A+E	\\
A+E	&	A+E	\\
A+F	&	A+F	\\
B+E	&	A+F	\\
B+F	&	A+F	\\
E+F	&	B+E	\\
A+B+E	&	B+E	\\
A+B+E	&	B+E	\\
A+B+E	&	B+F	\\
A+B+F	&	B+F	\\
A+B+F	&	B+F	\\
A+B+F	&	E+F	\\
A+E+F	&	E+F	\\
A+E+F	&	E+F	\\
A+E+F	&	A+B+E	\\
B+E+F	&	A+B+F	\\
B+E+F	&	A+E+F	\\
B+E+F	&	B+E+F	\\
A+B+E+F	&	A+B+E+F	\\
	&	A+B+E+F	\\
	&	A+B+E+F	\\
\hline
\caption{Pattern of unknown and known byproducts for SI $CD$}
\label{tab:27}

\end{longtable}
\end{center}

The query to $N_2$ should somehow accommodate these deviations without changing the structure. To construct a query to $N_{2}$, we use the query to $N_{1}$ and make appropriate modifications as discussed hereafter. For exposition, we take the case when $CD$ are the side information. We already know that the query to $N_{1}$ gives a symmetric known-unknown byproduct combinations for side information $AC$ and demand $G$. We pick the query submitted to $N_{1}$, and perform the transformation $A$ $\rightleftharpoons$ $D$, to generate a new query. Due to the transformation, it is clear that this new query gives a symmetric known-unknown byproduct combinations when $G$ is the demand and $CD$  are the side information.
 Since one bit of $A+B$ moves towards the unknown side from the known side, there is only one bit of $A+B$ available in known bits to retrieve $G$. But the new query after transformation uses 2 bits of $A+B$(bit number 39 and 43) to retrieve 2 bits of $G$ as seen in the third column of the Table \ref{tab:28}. This is rectified by swapping bit $G$ in bit number 39 with $F$, thereby retrieving the extra bit of $A+B+F$ required in the unknown whilst removing the necessity of second known bit of $A+B$. This increments and decrements the bit count (total number of bits of a message) of $F$ and $G$ respectively by 1. The extra bit of $A+B$ required is obtained by swapping bit $E$ in the bit number 19 with $B$ since all the two-tuple sum combinations except $A+B$ are unknown only once. This increments and decrements the bit count of $B$ and $E$ respectively by 1. Since the bit $A+F$ is unknown only once, the bit number 29, which is the second query for $A+F$ is changed to $B+D+G$ to use the extra known bit $B$ along with side information $D$ to retrieve $G$. This neutralises the differences occurred in bit count of $F$ and $G$ and also increments the bit count of $B$ and $D$ and decrements the bit count of $A$ and $C$ respectively by 1. $B$ has a net increment of 2 in the bit count. This increment of $D$ and decrement of $C$  is neutralised in bit number 56 by swapping $D$ with $C$. Now the four-tuple sum bit $A+B+E+F$ is known three times and is unknown only once. Therefore $B$ and side information $C$ are swapped in bit numbers 49 and 62. Bit number 49, which was initially querying a bit of $A+B+E+F$ will now query the extra unknown bit of $A+E+F$. Bit number 62 was supposed to be the second query for $G$ bit using $A+E+F$ but all the three-tuple sum bits have only one known bit. This $A+E+F$ gets replaced by the extra known bit of $A+B+E+F$. The extra unknown bit of three-tuple sum bit $B+E+F$ is retrieved by swapping $G$ with $F$ in bit number 40. This increments and decrements the bit count of $F$ and $G$ respectively by 1. Bit numbers 15 and 23 which queries the second bit $A$ and $B+E$ respectively are together used to retrieve the extra unknown bit of $A+B+E$ since $A$ and $B+E$ are unknown only once. Bit numbers 57-59 uses three-tuple sums $A+B+E, A+B+F$ and $B+E+F$ to retrieve $G$ for the second time but they are known only once. Therefore these bits are obtained by combining the bits $\{A,B+E\}$, $\{A,B+F\}$ and $\{B,E+F\}$ since $A$, $B$ and all the two-tuple sum blocks except $A+B$ are known one extra time. $B+D$ of bit number 38 is replaced with $A+E$ and $B$ of bit number 41 is replaced with $A$ so that the extra two-tuple sum bits, $A+E$ and $A+F$ will be used to retrieve G. This neutralises the bit count of $B$ and $E$ while increments the bit count of $A$ by 2 and decrements the bit count of $D$ by 1. Since bit count of $A$ was already lagging 1 bit behind, this step increases it by 2 bits to give a net increment of 1 for $A$. $C$ of bit number 33 is replaced with $B$ and $C+A$ of bit number 28 is replaced with $B+G$ so that the extra two-tuple sum bits, $B+E$ and $B+F$ will be used to retrieve $G$. This neutralises the bit count of $G$ and $A$ while decrements the bit count of $D$ by 1. This also increments and decrements the bit count of $B$ and $C$ respectively by 2. The $B$ from bit number 10 is swapped with side information $D$, while $D$ itself in bit number 1 is swapped with $G$. The $G$ bit and $B$ bit of bit numbers 14 and 2 are swapped with $C$ to obtain the second unknown bit of $F$ and thereby neutralising $B,C$ and $G$. $B$ is swapped with $E$ in both bit numbers 6 and 9 to retrieve extra unknown bit $E$ and only unknown bit of $E+F$. This increments and decrements the bit count of $E$ and $B$ respectively by 2. $E$ in bit numbers 21 and 31 are swapped with $G$ and $B$ respectively and $F$ in bit number 31 is swapped with $D$ to retrieve extra unknown bit of $F$ and only unknown bit of $B$. This neutralises $D,E$ and $F$ while increments $B$ and $G$ one time with $B$ having a net decrement of 1 bit. $G$ of  bit number 13 is swapped with $A$ to retrieve the only known bit of $A+E$ while $A$ of bit number 30 is swapped with the last known bit of $B$ to retrieve $G$ which neutralises the bit count of $B$ and $G$ and achieving the uniform distribution of bits while keeping the structure of query at $N_2$ maintained as that of $N_1$. This sequence of operations from taking a copy of the query to $N_{1}$, the transformation of $A$ $\rightleftharpoons$ $D$, and the final exchange in bit positions are displayed in the last three columns of Table \ref{tab:28}. Thus, the last column of this table forms the query to $N_{2}$ when $CD$ are the side information. Of course, swapping of the known-unknown byproducts indexes is required wherever they are symmetric. 
\begin{center}							
\begin{longtable}{ |c|c|c|c| }							
\hline							
BIT NUMBER&$N_1$&$A$ $\rightleftharpoons$ $D$&$N_2$\\						
\hline							
\textcolor{red}1	&	\textcolor{red}A	&	\textcolor{red}D	&	\textcolor{red}G	\\
\textcolor{red}2	&	\textcolor{red}{A+B}	&	\textcolor{red}{D+B}	&	\textcolor{red}{D+C}	\\
3	&	A+D	&		D+A	&	D+A\\
4	&	A+E	&	D+E	&	D+E\\
5	&	A+F	&	D+F	&	D+F		\\
\textcolor{red}6	&	\textcolor{red}{B+C}	&	\textcolor{red}{B+C}	&	\textcolor{red}{E+C}	\\
7	&	B+D&		B+A	&	B+A		\\
8	&	B+E	&B+E	&	B+E\\
\textcolor{red}{9}	&	\textcolor{red}{B+F}	&\textcolor{red}{B+F}	&	\textcolor{red}{E+F}\\
\textcolor{red}{10}	&	\textcolor{red}{B+G}	&	\textcolor{red}{B+G}	&	\textcolor{red}{D+G}	\\
11	&	C+G&		C+G	&	C+G\\
12 &	D+G&		A+G	&	A+G\\
\textcolor{red}{13}	&	\textcolor{red}{E+G}	&	\textcolor{red}{E+G}	&	\textcolor{red}{E+A}	\\
\textcolor{red}{14}	&	\textcolor{red}{F+G}	&	\textcolor{red}{F+G}	&	\textcolor{red}{F+C}	\\
15	&	A+C+D	&	D+C+A	&	D+C+A	\\
16	&A+C+E	&D+C+E	&	D+C+E	\\
17	&	A+C+F	&D+C+F	&	D+C+F	\\
18	&	A+C+G	&D+C+G	&	D+C+G\\
\textcolor{red}{19}	&	\textcolor{red}{A+D+E}	&	\textcolor{red}{D+A+E}	&	\textcolor{red}{D+A+B}	\\
20	&	A+D+F	&D+A+F	&	D+A+F	\\
\textcolor{red}{21}	&	\textcolor{red}{A+E+F}	&	\textcolor{red}{D+E+F}	&	\textcolor{red}{D+G+F}	\\
22	&	B+C+D	&B+C+A	&	B+C+A	\\
23	&	B+C+E	&		B+C+E	&		B+C+E	\\
24	&	B+C+F	&	B+C+F	&	B+C+F	\\
\textcolor{black}{25}	&	\textcolor{black}{B+D+E}	&	\textcolor{black}{B+A+E}	&	\textcolor{black}{B+A+E}	\\
26	&	B+D+F	&	B+D+F	&	B+D+F	\\
\textcolor{black}{27}	&	\textcolor{black}{B+E+F}	&	\textcolor{black}{B+E+F}	&	\textcolor{black}{B+E+F}\\
\textcolor{red}{28}	&	\textcolor{red}{C+D+E}	&	\textcolor{red}{C+A+E}	&	\textcolor{red}{B+G+E}		\\
\textcolor{red}{29}	&	\textcolor{red}{C+D+F}	&	\textcolor{red}{C+A+F}	&	\textcolor{red}{B+D+G}	\\
\textcolor{red}{30}	&	\textcolor{red}{C+D+G}	&	\textcolor{red}{C+A+G}	&	\textcolor{red}{C+B+G}	\\
\textcolor{red}{31}	&	\textcolor{red}{C+E+F}	&	\textcolor{red}{C+E+F}	&	\textcolor{red}{C+B+D}	\\
32	&	C+E+G	&	C+E+G	&	C+E+G	\\
\textcolor{red}{33}	&	\textcolor{red}{C+F+G}	&	\textcolor{red}{C+F+G}	&	\textcolor{red}{B+F+G}	\\
34	&	D+E+F	&	A+E+F	&	A+E+F	\\
35	&	D+E+G	&	A+E+G	&	A+E+G		\\
36	&	D+F+G	&	A+F+G	&	A+F+G		\\
37	&	E+F+G	&	E+F+G	&	E+F+G	\\
\textcolor{red}{38}	&	\textcolor{red}{A+B+C+G}	&	\textcolor{red}{D+B+C+G}	&	\textcolor{red}{A+E+C+G}	\\
\textcolor{red}{39}	&	\textcolor{red}{A+B+D+G}	&	\textcolor{red}{D+B+A+G}	&	\textcolor{red}{A+B+F+D}	\\
\textcolor{red}{40}	&	\textcolor{red}{A+B+E+G}	&	\textcolor{red}{D+B+E+G}	&	\textcolor{red}{D+B+E+F}	\\
\textcolor{red}{41}	&	\textcolor{red}{A+B+F+G}	&	\textcolor{red}{D+B+F+G}	&	\textcolor{red}{D+A+F+G}	\\
42	&	C+D+E+F	&	C+A+E+F	&	C+A+E+F	\\
43	&	A+B+C+D+E	&	D+B+C+A+E	&	D+B+C+A+E	\\
44	&	A+B+C+D+F	&	D+B+C+A+F	&	D+B+C+A+F	\\
45	&	A+B+C+D+G	&	D+B+C+A+G	&	D+B+C+A+G	\\
46	&	A+B+C+E+F	&	D+B+C+E+F	&	D+B+C+E+F	\\
47	&	A+B+C+E+G	&	D+B+C+E+G	&	D+B+C+E+G	\\
48	&	A+B+C+F+G	&	D+B+C+F+G	&	D+B+C+F+G	\\
\textcolor{red}{49}	&	\textcolor{red}{A+B+D+E+F}	&	\textcolor{red}{D+B+A+E+F}	&	\textcolor{red}{D+C+A+E+F}	\\
50	&	A+B+D+E+G	&	D+B+A+E+G	&	D+B+A+E+G	\\
51	&	A+B+D+F+G	&	D+B+A+F+G	&	D+B+A+F+G	\\
52	&	A+B+E+F+G	&	D+B+E+F+G	&	D+B+E+F+G	\\
53	&	A+C+D+E+G	&	D+C+A+E+G	&	D+C+A+E+G	\\
54	&	A+C+D+F+G	&	D+C+A+F+G	&	D+C+A+F+G	\\
55	&	A+C+E+F+G	&	D+C+E+F+G	&	D+C+E+F+G	\\
\textcolor{red}{56}	&	\textcolor{red}{A+D+E+F+G}	&	\textcolor{red}{D+A+E+F+G}	&	\textcolor{red}{C+A+E+F+G}	\\
57	&	B+C+D+E+G	&	B+C+A+E+G	&	B+C+A+E+G	\\
58	&	B+C+D+F+G	&	B+C+A+F+G	&	B+C+A+F+G	\\
59	&	B+C+E+F+G	&	B+C+E+F+G	&	B+C+E+F+G	\\
60	&	B+D+E+F+G	&	B+A+E+F+G	&	B+A+E+F+G	\\
61	&	A+B+C+D+E+F	&	D+B+C+A+E+F	&	D+B+C+A+E+F	\\
\textcolor{red}{62}	&	\textcolor{red}{A+C+D+E+F+G}	&	\textcolor{red}{D+C+A+E+F+G}	&	\textcolor{red}{D+B+A+E+F+G}	\\
63	&	B+C+D+E+F+G	&	B+C+A+E+F+G	&	B+C+A+E+F+G	\\
\hline		
\caption{Final query to $N_2$ given in the fourth column.}
\label{tab:28}
\end{longtable}							
\end{center}

In general, when handling the other cases of $\{CE,CF\}$ as the side information, the procedure for obtaining query for $N_2$ is similar to that when $CD$ are the side information. We pick the query submitted to $N_1$ and perform the transformation between $A$ with $D$. The modifications made to the new query after transformation when $CD$ were the side information can be reproduced to retrieve $G$ from $N_2$.

\subsubsection{One of $\{DE,DF,EF\}$ is the side information and $G$ is the demand}
\label{sec:g25}

Compared to the previous case (Section \ref{sec:g24}), this case has one more singleton bit of $A$ going to the known side from the unknown side and one bit of $A+C$ going from the known side to the unknown side for all side information from $\{DE,DF,EF\}$. A minor modification to the previous code for $N_2$ seen in Table \ref{tab:28} can incorporate the change required. Since the fourth column of Table \ref{tab:28} gave the query to $N_2$ for demand $G$ and side information $CD$, swap $CD$ with the desired side information from  $\{DE,DF,EF\}$ to obtain a new query.  Since one more $A$ goes to the known side, all 4 bits of $A$ are known. Therefore a side information in the bit number 15 in the fourth column of Table \ref{tab:28} which queries the only unknown $A$ can be replaced with $C$ to make it retrieve $A+C$. Bit number 38 which retrieves $G$ using $A+C$ can be now used to retrieve $G$ using the fourth known $A$ bit by swapping $C$ with the side information swapped before. Therefore all bits of $G$ can be retrieved from both databases when one of $\{DE,DF,EF\}$ is the side information without altering the structure of query from $N_1$.


\subsection{G as a byproduct}

\label{sec:g3}
G can be byproduct in (the demand can come from $\{A,B,C,D,E,F\}$ in $K-1 \choose 1$ = $6 \choose 1$ = 6) $\times$ (the 2 side information messages can come from $\{A,B,C,D,E,F\}$ - $Demand$ in $K-2 \choose 2$ = $5 \choose 2$ = 10) = 6$\times$10 = 60 ways. This is subdivided into 4 different cases.
\subsubsection{$A$ is the demand}
    When  $A$ is the demand and $BC$ are the side information, the known and unknown byproduct combinations are symmetric. The demand can be retrieved similar to the operation mentioned in step 8 of the code construction in Section \ref{sec:3}. Now when one of $\{BD,BE,BF\}$ are the side information there is a small difference from symmetry. In the unknown side one singleton bit of $C$ and $G$ gets removed(doesn't go to the known side) and one bit of $C+G$ gets added to unknown(doesn't get removed from known). The structure of code remains the same as that of $N_1$  since the $C$ and $G$ of the extra unknown $C+G$ bit can be obtained from the individual querying itself since one singleton bit of $C$ and $G$ is removed from unknown. Now when one of $\{CD,CE,CF\}$ are the side information, the known and unknown byproduct combinations are similar to the case when $G$ was the demand and one of $\{CD,CE,CF\}$ were the side information (Section \ref{sec:g24}). A simple transformation of $A$ $\rightleftharpoons$ $G$ in the code structure provided in Section \ref{sec:g24} will provide the necessary query to $N_2$. Now when one of $\{DE,DF,EF\}$ is the side information, the known and unknown byproduct combinations are similar to the case of $\{CD,CE,CF\}$ with a a small difference. The difference is same as $\{BD,BE,BF\}$ case, in the unknown side one singleton bit of $C$ and $G$ gets removed(doesn't go to the known side) and one bit of $C+G$ gets added to unknown(doesn't get removed from known). The structure of code remains the same as $\{CD,CE,CF\}$ since the $C$ and $G$ of the extra unknown $C+G$ bit can be obtained from the individual querying itself since one singleton bit of $C$ and $G$ is removed from unknown. This completes all side information cases for $A$ as a demand and $G$ as one of the byproducts. Therefore all cases with $A$ as a demand is retrievable.

\subsubsection{$B$ is the demand}
    \label{sec:g32}
    When $B$ is the demand, the cases are similar to $A$ as the demand except for the side information set $\{DE,DF,EF\}$. When $B$ is the demand and $AC$ are the side information, the known and unknown byproduct combinations are symmetric. The demand can be retrieved similar to the operation mentioned in step 8 of the code construction in Section \ref{sec:3}. Now when one of $\{AD,AE,AF\}$ are the side information there is a small difference from symmetry. In the unknown side one singleton bit of $C$ and $G$ gets removed(doesn't go to the known side) and one bit of $C+G$ gets added to unknown(doesn't get removed from known). The structure of code remains the same as that of $N_1$ since the $C$ and $G$ of the extra unknown $C+G$ bit can be obtained from the individual querying itself since one singleton bit of $C$ and $G$ is removed from unknown. Now when one of $\{CD,CE,CF\}$ are the side information, the known and unknown byproduct combinations are similar to the case when $G$ was the demand and one of $\{CD,CE,CF\}$ were the side information (Section \ref{sec:g24}). A simple transformation of $B$ $\rightleftharpoons$ $G$ in the code structure provided in Section \ref{sec:g24} will provide the necessary query to $N_2$. Now when one of $\{DE,DF,EF\}$ is the side information, the known and unknown byproduct combinations are similar to the case when $G$ was the demand and one of $\{DE,DF,EF\}$ were the side information (Section \ref{sec:g25}) with a small difference. The unknown bit $A+C$ is replaced with just $A$ while the known bit $A+G$ is replaced with $A+C+G$. When $G$ was the demand and one of $\{CD,CE,CF\}$ were the side information, for obtaining second known bit of $A+B+E$  we had to combine individual bits of $A$ and $B+E$ as explained in Section \ref{sec:g24}. Similar operation was performed for $A+B+C$ in Section \ref{sec:g25}. Since second $A+C+G$(which is analogous of $A+B+C$ from Section \ref{sec:g25}) bit is available as known, we can use it directly to query demand $B$ thereby freeing individual bits of $A$ and $C+G$. Now this free known bit $C+G$ is used in place of retrieving $B$ with singleton $G$. This frees a singleton $G$ and this along with the singleton $A$ freed before is used to retrieve $B$ with bit $A+G$ (since the only known $A+G$ was converted to $A+C+G$). Finally the query bit that retrieves $A+C$ is now used to retrieve the new unknown $A$ bit. This completes all side information cases for $B$ as a demand and $G$ as one of the byproducts. Therefore all cases with $B$ as a demand is retrievable.
    \subsubsection{$C$ is the demand}
    \label{sec:g33}
    When $C$ is the demand and one of $\{AB,AD,AE,AF\}$ are the side information, the known and unknown byproduct combinations are similar to the case when $G$ was the demand and one of $\{AB,AD,AE,AF\}$ were the side information (Section \ref{sec:g22}). The query for $N_2$ can be obtained similar to that case with a transformation of $C$ $\rightleftharpoons$ $G$. When one of $\{BD,BE,BF\}$ are the side information, the known and unknown byproduct combinations are similar to the case when $G$ was the demand and one of $\{BD,BE,BF\}$ were the side information (Section \ref{sec:g23}). The query for $N_2$ can be obtained similar to that case with a transformation of $C$ $\rightleftharpoons$ $G$. When one of $\{DE,DF,EF\}$ are the side information, the known and unknown byproduct combinations is similar to the one when $B$ was the demand and one of $\{DE,DF,EF\}$ were the side information (Section \ref{sec:g32}) with some minor differences. In the unknown side, one bit of $A+G$ is removed while one bit of $A$ and 2 bits of $G$ are included. In the known side 2 bits of $B+G$ are replaced by just $B$ while one bit of $A+B$ is replaced by $A+B+G$. For instance, when $EF$ are the side information,the known and unknown byproduct combinations are as shown in the first two columns of Table \ref{tab:29}.
    \begin{center}
\begin{longtable}{ |c|c| }
\hline
UNKNOWN&KNOWN\\
\hline
    A	&	A	\\
A	&	A	\\
B	&	A	\\
B	&	A	\\
B	&	B	\\
D	&	B	\\
D	&	B	\\
D	&	G	\\
G	&	G	\\
G	&	G	\\
G	&	D	\\
A+G	&	A+B	\\
A+G	&	B+G	\\
B+G	&	D+G	\\
A+B	&	D+G	\\
D+G	&	D+G	\\
A+D	&	A+D	\\
B+D	&	A+D	\\
A+B+G	&	A+D	\\
A+B+G	&	B+D	\\
A+B+G	&	B+D	\\
A+D+G	&	B+D	\\
A+D+G	&	A+B+G	\\
A+D+G	&	A+B+G	\\
B+D+G	&	A+B+G	\\
B+D+G	&	A+D+G	\\
B+D+G	&	B+D+G	\\
A+B+D	&	A+B+D	\\
A+B+D	&	A+B+D+G	\\
A+B+D	&	A+B+D+G	\\
A+B+D+G	&	A+B+D+G	\\
\hline
\caption{Pattern of unknown and known byproducts for SI $EF$}
\label{tab:29}
    \end{longtable}
    \end{center}
    The query to $N_2$ should somehow accommodate these deviations without changing the structure. For exposition, we take the case when $EF$ are the side information. To construct a query to $N_{2}$, we use the query to $N_{2}$ from Table \ref{tab:28} and make appropriate modifications as discussed hereafter. To this query we perform the transformation $CD$ $\rightleftharpoons$ $EF$, to generate a new query. To this query we perform the modifications mentioned as in the case of $G$ as the demand and one of $\{DE,DF,EF\}$ as the side information to obtain a new query. To this query we perform a transformation $B$ $\rightleftharpoons$ $G$ to obtain a new query. To this query we perform the modifications mentioned as in the case of $B$ as the demand and one of $\{DE,DF,EF\}$ as the side information to obtain a new query. Note that this query is exactly the query to $N_2$ for the case of $B$ as the demand and $EF$ are the side information. To this query we perform a transformation of $B$ $\rightleftharpoons$ $C$ as seen in third column of Table \ref{tab:30}. Note that the third column of Table \ref{tab:30} is not obtained by direct transformation of $B$ $\rightleftharpoons$ $C$ in $N_1$. Rather, it is obtain after the execution of various steps as discussed earlier. Since one bit of $A+G$ is removed from the unknown, bit number 7 which was initially querying $A+G$ can now be used to query one extra bit of $G$. Bit number 2 which was just a combination of side information $E+F$ can be used to query second extra bit of $G$. Bit number 8 and 28 are modified to query $C$ with two extra known singleton $B$ bits since one $B+G$ bit in known set is removed. Since bit number 8 which queried $B+G$ which was required to obtain third bit of unknown $A+B+G$ is modified, the third bit of $A+B+G$ is queried by modifying bit number 38. This modification removes the retrieval of one bit of demand using $A+D$. This is neutralised by modifying bit number 41 to remove query of demand with $B+G$ (since 2 bits of $B+G$ was removed from known set) and replace it with $A+D$. Since this $A+B+G$ is queried directly, $A$ retrieved in bit number 11 which was supposed to be the $A$ in $A+B+G$ can be now used to retrieve the extra singleton bit $A$ which came to the unknown side. Finally bit number 53 which was the bit that used $A+B$ to query $C$ can now be used to query $A+B+D+G$(since one bit of $A+C$ gets removed from known) while bit number 61 which was initially supposed to query $A+B+D+G$ is now used to retrieve $C$ with the extra $A+B+G$ bit that is available in the known set. Query to $N_1$, the transformation of $B$ $\rightleftharpoons$ $C$ to query of $N_2$ for the case of $B$ as the demand and $EF$ are the side information, and the final exchange in bit positions are displayed in the last three columns of Table \ref{tab:30}. Thus, the last column of this table forms the query to $N_{2}$ when $EF$ are the side information and $C$ is the demand.
\begin{center}							
\begin{longtable}{ |c|c|c|c| }							
\hline							
BIT NUMBER&$N_1$&$B$ $\rightleftharpoons$ $C$&$N_2$\\						
\hline			
1	&	A	&	C	&	C	\\
\textcolor{red}{2}	&	\textcolor{red}{A+B}	&	\textcolor{red}{F+E}	&	\textcolor{red}{F+G}	\\
3	&	A+D	&	F+A	&	F+A	\\
4	&	A+E	&	F+B	&	F+B	\\
5	&	A+F	&	F+D	&	F+D	\\
6	&	B+C	&	B+E	&	B+E	\\
\textcolor{red}{7}	&	\textcolor{red}{B+D}	&	\textcolor{red}{G+A}	&	\textcolor{red}{G+E}	\\
\textcolor{red}{8}	&	\textcolor{red}{B+E}	&	\textcolor{red}{G+B}	&	\textcolor{red}{C+B}	\\
9	&	B+F	&	B+D	&	B+D	\\
10	&	B+G	&	F+C	&	F+C	\\
11	&	C+G	&	E+C	&	E+C	\\
12	&	D+G	&	A+C	&	A+C	\\
13	&	E+G	&	E+A	&	E+A	\\
14	&	F+G	&	D+E	&	D+E	\\
15	&	A+C+D	&	E+B+A	&	E+B+A	\\
16	&	A+C+E	&	F+E+B	&	F+E+B	\\
17	&	A+C+F	&	F+E+D	&	F+E+D	\\
18	&	A+C+G	&	F+E+C	&	F+E+C	\\
19	&	A+D+E	&	F+A+G	&	F+A+G	\\
20	&	A+D+F	&	F+A+D	&	F+A+D	\\
21	&	A+E+F	&	F+C+D	&	F+C+D	\\
22	&	B+C+D	&	G+E+A	&	G+E+A	\\
23	&	B+C+E	&	G+E+B	&	G+E+B	\\
24	&	B+C+F	&	G+E+D	&	G+E+D	\\
25	&	B+D+E	&	G+A+B	&	G+A+B	\\
26	&	B+D+F	&	G+F+D	&	G+F+D	\\
27	&	B+E+F	&	G+B+D	&	G+B+D	\\
\textcolor{red}{28}	&	\textcolor{red}{C+D+E}	&	\textcolor{red}{G+C+B}	&	\textcolor{red}{F+C+B}	\\
29	&	C+D+F	&	G+F+C	&	G+F+C	\\
30	&	C+D+G	&	E+G+C	&	E+G+C	\\
31	&	C+E+F	&	E+G+F	&	E+G+F	\\
32	&	C+E+G	&	E+B+C	&	E+B+C	\\
33	&	C+F+G	&	G+D+C	&	G+D+C	\\
34	&	D+E+F	&	A+B+D	&	A+B+D	\\
35	&	D+E+G	&	A+B+C	&	A+B+C	\\
36	&	D+F+G	&	A+D+C	&	A+D+C	\\
37	&	E+F+G	&	B+D+C	&	B+D+C	\\
\textcolor{red}{38}	&	\textcolor{red}{A+B+C+G}	&	\textcolor{red}{B+G+F+C}	&	\textcolor{red}{A+D+E+C}	\\
39	&	A+B+D+G	&	A+G+D+F	&	A+G+D+F	\\
40	&	A+B+E+G	&	F+G+B+D	&	F+G+B+D	\\
\textcolor{red}{41}	&	\textcolor{red}{A+B+F+G}	&	\textcolor{red}{F+A+D+C}	&\textcolor{red}{F+A+B+G}	\\
42	&	C+D+E+F	&	E+A+B+D	&	E+A+B+D	\\
43	&	A+B+C+D+E	&	F+G+E+A+B	&	F+G+E+A+B	\\
44	&	A+B+C+D+F	&	F+G+E+A+D	&	F+G+E+A+D	\\
45	&	A+B+C+D+G	&	F+G+E+A+C	&	F+G+E+A+C	\\
46	&	A+B+C+E+F	&	F+G+E+B+D	&	F+G+E+B+D	\\
47	&	A+B+C+E+G	&	F+G+E+B+C	&	F+G+E+B+C	\\
48	&	A+B+C+F+G	&	F+G+E+D+C	&	F+G+E+D+C	\\
49	&	A+B+D+E+F	&	F+E+A+B+D	&	F+E+A+B+D	\\
50	&	A+B+D+E+G	&	F+G+A+B+C	&	F+G+A+B+C	\\
51	&	A+B+D+F+G	&	F+G+A+D+C	&	F+G+A+D+C	\\
52	&	A+B+E+F+G	&	F+G+B+D+C	&	F+G+B+D+C	\\
\textcolor{red}{53}	&	\textcolor{red}{A+C+D+E+G}	&	\textcolor{red}{F+E+A+B+C}	&	\textcolor{red}{A+G+B+D+F}	\\
54	&	A+C+D+F+G	&	F+E+A+D+C	&	F+E+A+D+C	\\
55	&	A+C+E+F+G	&	F+E+B+D+C	&	F+E+B+D+C	\\
56	&	A+D+E+F+G	&	E+A+B+D+C	&	E+A+B+D+C	\\
57	&	B+C+D+E+G	&	G+E+A+B+C	&	G+E+A+B+C	\\
58	&	B+C+D+F+G	&	G+E+A+D+C	&	G+E+A+D+C	\\
59	&	B+C+E+F+G	&	G+E+B+D+C	&	G+E+B+D+C	\\
60	&	B+D+E+F+G	&	G+A+B+D+C	&	G+A+B+D+C	\\
\textcolor{red}{61}	&	\textcolor{red}{A+B+C+D+E+F}	&	\textcolor{red}{F+G+E+A+B+D}	&	\textcolor{red}{A+B+G+C+E+F}	\\
62	&	A+C+D+E+F+G	&	F+G+A+B+D+C	&	F+G+A+B+D+C	\\
63	&	B+C+D+E+F+G	&	G+E+A+B+D+C	&	G+E+A+B+D+C	\\
\hline				
\caption{Final query to $N_2$ given in the fourth column.}
\label{tab:30}
\end{longtable}							
\end{center}	
In general, when handling the other cases of $\{DE,DF\}$ as the side information, the procedure for obtaining query for $N_2$ is similar to that when $EF$ are the side information. We pick the query submitted to $N_2$ for the case of $B$ as the demand and  $\{DE,DF\}$ as the side information respectively and perform the transformation between $B$ with $C$. The modifications made to the new query after transformation when $EF$ were the side information can be reproduced to retrieve $B$ from $N_2$.

\label{sec:g34}
\subsubsection{One of $\{D,E,F\}$ is the demand}
    \label{sec:g34}
When one of $\{D,E,F\}$ is the demand, the known and unknown byproduct combinations are symmetric if either $A$ or $B$ is one of the side information. The demand can be retrieved similar to the operation mentioned in step 8 of the code construction in Section \ref{sec:3}. For other side information pairs which doesn't have either $A$ or $B$ in side information,the known and unknown byproduct combinations are same as the previous case where $C$ was the demand and one of $\{DE,DF,EF\}$ were side information with two exceptions. One bit of $A$ goes to the unknown side while one bit of $A+G$ comes to the known side. For exposition let us consider $D$ as the demand. To obtain the query for $N_2$, perform the transformation of $C$ $\rightleftharpoons$ $D$ to the query for $N_2$ in Table \ref{tab:30}. To this new query use any bit that queries $A+G$ to query $A$. Use any bit that retrieves $D$ by using $A$ to retrieve $D$ by using $A+G$. \par In general, when handling the other cases of $\{E,F\}$ as the demand, pick the query submitted to $N_2$ for the case of $C$ as the demand and $\{DE,DF,EF\}$ as the side information respectively and perform the transformation between $C$ with demand. The modifications made to the new query after transformation when $D$ was the demand can be reproduced to retrieve demand from $N_2$. This completes all side information cases when one of $\{D,E,F\}$ is the demand and $G$ is one of the byproducts. Therefore all cases with one of $\{D,E,F\}$ as the demand are retrievable.

 \section{Discussion and Directions for Future Work}
 \label{sec:5}
 
 In this work, we have presented the first XOR-based code construction for a PIR-PSI setting involving $N = 2$ databases and $M = 2$ side information. Although our code construction marginally falls short of the capacity of PIR-PSI setting, it offers substantial reduction in the decoding complexity when compared to the MDS counterpart (which achieves the capacity). Importantly, we have shown that our codes provide rates strictly higher than the capacity of PIR schemes. This implies that our codes are applicable when multiple files have to be downloaded by a user at different time-instants from the two databases. We believe that this work can be extended in one of the following directions, namely: (i) How to construct XOR-based PSR-PSI codes for $N = 2$ non-colluding databases with arbitrary values of $M$ side information messages? (ii) How to construct XOR-based PSR-PSI codes for arbitrary values of $N$ and $M$? and finally, (iii) Do XOR-based PIR-PSI codes exist that achieve the capacity of the PSR-PSI setting?


\begin{thebibliography}{1}
\bibitem{chor}
B.Chor, O.Goldreich, E.Kushilevitz, and M.Sudan, ``Private information retrieval," \emph{36th Annual Symposium on Foundations of Computer Science}, 1995, pp. 41–50.

\bibitem{sunjafar}
Hua Sun and Syed Ali Jafar, ``The Capacity of Private Information Retrieval," \emph{IEEE Transactions on Information Theory}, vol. 63, no. 7, pp. 4075-4088, Jul. 2017.

\bibitem{banawan}
K.Banawan and S.Ulukus,  ``The capacity of private information  retrieval from coded databases," \emph{IEEE Transactions on Information Theory,} vol. 64, no. 3, pp. 1945-1956, Mar. 2018.

\bibitem{tajeddine}
R. Tajeddine,  O.W. Gnilke, and S. El Rouayheb, ``Private  information retrieval  from  mds  coded  data  in  distributed  storage  systems," \emph{IEEE Transactions on Information Theory}, vol. 64, no. 11, pp. 7081-7093, Nov. 2018.

\bibitem{sunjafar2}
H. Sun and S.A. Jafar, ``The  capacity  of  robust  private  information retrieval  with  colluding  databases," \emph{IEEE Transactions on Information Theory}, vol. 64, no. 4, pp. 2361-2370, Apr. 2018.
    
\bibitem{sunjafar3}
H. Sun and S. A. Jafar, ``The Capacity of Symmetric Private Information Retrieval," \emph{IEEE Transactions on Information Theory}, vol. 65, no. 1, pp. 322-329, Jan. 2019.

\bibitem{tandon}
R. Tandon, ``The capacity of cache aided private information retrieval," \emph{2017 55th Annual Allerton Conference on Communication, Control, and Computing (Allerton)}, 2017, pp. 1078-1082.

\bibitem{kadhe}
Swanand Kadhe, Brenden Garcia, Anoosheh Heidarzadeh, Salim El Rouayheb, and Alex Sprintson, ``Private information retrieval with side information: The single server case," \emph{2017 55th Annual Allerton Conference on Communication, Control, and Computing (Allerton)}, 2017, pp. 1099-1106.

\bibitem{Kazemi}
A. Heidarzadeh, F. Kazemi and A. Sprintson, ``Capacity of Single-Server Single-Message Private Information Retrieval with Coded Side Information," \emph{2018 IEEE Information Theory Workshop (ITW),} 2018, pp. 1-5.

\bibitem{suli}
S. Li and M. Gastpar, ``Converse for Multi-Server Single-Message PIR with Side Information," \emph{2020 54th Annual Conference on Information Sciences and Systems (CISS),} 2020, pp. 1-6.

\bibitem{shariat}
S. P. Shariatpanahi, M. J. Siavoshani and M. A. Maddah-Ali, ``Multi-Message Private Information Retrieval with Private Side Information," \emph{2018 IEEE Information Theory Workshop (ITW),} 2018, pp. 1-5.

\bibitem{zchen}
Z. Chen, Z. Wang and S. A. Jafar, ``The Capacity of T-Private Information Retrieval With Private Side Information," \emph{IEEE Transactions on Information Theory}, vol. 66, no. 8, pp. 4761-4773, Aug.2020.
\end{thebibliography}
\end{document}